\documentclass[useAMS,usenatbib]{mnras}
\pdfoutput=1
\newcommand{\msun}{{\ensuremath{{\rm M}_{\odot}}}}
\usepackage{rotating, graphicx}
\usepackage{color}
\usepackage{txfonts}
\usepackage{amssymb} 
\usepackage{lscape}
\usepackage{float}
\usepackage{verbatim} 

\title[]{A chronicle of galaxy mass assembly in the EAGLE simulation}
\author[Qu et al.]{Yan Qu$^{1}$\thanks{E-mail:quyan@nao.cas.cn},  John C. Helly$^{2}$, 
Richard G. Bower$^{2}$, Tom Theuns$^{2}$, Robert A. Crain$^{3}$, \and 
Carlos S. Frenk$^{2}$, Michelle Furlong$^{2}$, Stuart McAlpine$^{2}$, Matthieu Schaller$^{2}$, \and 
Joop Schaye$^{4}$, Simon D. M. White$^{5}$\\
$^{1}$National Astronomical Observatories, Chinese Academy of Sciences, 20A Datun Road, Chaoyang, Beijing 10012, China\\
$^{2}$Institute of Computational Cosmology, Durham University, South Road, Durham DH1 3LE, UK\\
$^{3}$Astrophysics Research Institute, Liverpool John Moores University, 146 Brownlow Hill, Liverpool L3 5RF, UK\\
$^{4}$Leiden Observatory, Leiden University, Postbus 9513, NL-2300 RA Leiden, the Netherlands\\
$^{5}$Max-Planck-Institut f$\ddot{u}$r Astrophysik, Karl-Schwarzschild-Strae 1, D-85741 Garching, Germany}

\begin{document}
\date{Accepted   Received}
\maketitle

\begin{abstract}
We analyse the mass assembly of central galaxies in the EAGLE hydrodynamical simulations. 
We build merger trees to connect galaxies to their progenitors at different redshifts and characterize 
their assembly histories by focusing on the time when half of the galaxy stellar mass was assembled 
into the main progenitor. We show that galaxies with stellar mass $M_*<10^{10.5}\msun$ 
assemble most of their stellar mass through star formation in the main progenitor (`in-situ' star formation). 
This can be understood as a consequence of the steep rise in star formation efficiency with halo mass for these 
galaxies. For more massive galaxies, however, an increasing fraction of their stellar mass is formed outside the main progenitor and 
subsequently accreted. Consequently, while for low-mass galaxies the assembly time is close to the stellar 
formation time, the stars in high-mass galaxies typically formed long before half of the present-day stellar mass was 
assembled into a single object, giving rise to the observed anti-hierarchical downsizing trend. 
In a typical present-day $M_*\geq10^{11}\msun$ galaxy, around $20\%$ of the 
stellar mass has an external origin. This fraction decreases with increasing redshift. Bearing in mind that 
mergers only make an important contribution to the stellar mass growth of massive galaxies, 
we find that the dominant contribution comes from mergers with galaxies of mass greater than one tenth 
of the main progenitor\rq{}s mass. The galaxy merger fraction derived from our simulations agrees with 
recent observational estimates.
\end{abstract}


\section{Introduction} 
In the Cold Dark Matter cosmological model, the growth of dark matter haloes is largely 
self-similar, with larger haloes being formed more recently 
than their low-mass counterparts. The formation and assembly of galaxies are, however, 
much more complex. Feedback from massive stars and the formation of black holes generates 
a strongly non-linear relationship between the masses of dark matter haloes and those of the 
galaxies they host. For low-mass haloes (with mass $\lesssim 10^{11.5} \msun$) the stellar mass 
increases rapidly, with a slope of $\sim2$, but in higher-mass haloes, the stellar mass of the main 
(or `central') galaxy increases much more slowly than the halo mass, with a slope of $\sim0.5$ 
\citep[e.g.][]{bensonEtal03, behrooziWC13,mosterNW13}. The mass assembly of galaxies 
will therefore be quite different from those of their parent haloes. 
Establishing how galaxies assemble their stars over cosmic time is then central to understanding 
galaxy formation and evolution. 

One question we need to answer, is the relative importance of the growth of galaxies via 
internal ongoing star formation (`in-situ'), in comparison to the mass contributions of 
external processes \citep[e.g.][]{guoW08, ZolotovEtal09, oserEtal10, fontEtal11, McCarthyEtal12, PillepichMM15}. 
These external processes, can be further divided to distinguish between 
the mass growth due to mergers with galaxies of comparable mass (`major mergers'), and 
the mass gained from much smaller galaxies (`minor mergers') or barely resolved systems and 
diffuse mass (`accretion'). While major mergers can rapidly increase a galaxy\rq{}s stellar mass, 
minor mergers are much more common \citep[e.g.][]{hopkinsEtal08, parryEF09}.  

To evaluate the relative importance of mergers to galaxy assembly, we need to know their 
merging histories. From an observational perspective, counts of close galaxy pairs 
\citep[e.g.][]{williamsQF11,manZT14}, or galaxies with disturbed morphologies 
\citep[e.g.][]{lotzEtal08,conseliceYB09,LopezSanjuanEtal11, stottEtal13}, provide a census of galaxy mergers. 
These values can be further converted into galaxy merger rates through the use of a 
merger timescale \citep[e.g.][]{kitzbichlerW08}. Unfortunately, those methods have 
their own limitations: galaxies in close-pairs may not be physically related, and may be 
chance line-of-sight superpositions; Morphological disturbances are not unique to 
galaxy mergers. For example, clumpy star formation driven by gravitational instability can 
also foster the formation of galaxies with irregular morphologies \citep{lotzEtal08}. 
In addition, these methods are sensitive to the merger stage and the mass ratio of the 
merging galaxies. Due to these limitations, the scatter between merger rate measurements is 
large, and it is difficult to make a reliable assessment of the complementary contribution of 
mergers to galaxy growth. Recently, deep surveys have begun to shed more light on the 
galaxy merger rate at high redshifts \citep[e.g.][]{manZT14}. Even so, the evolution of 
the merger rate remains controversial. An alternative approach is to extract the merger rates of 
galaxies from a model that reproduces the observed abundance of galaxies (and their 
distribution in mass), and its evolution with redshift, in a full cosmological context.

In the hierarchical structure formation scenario, the assembly of galaxies is believed to be 
closely related to the formation histories of their parent haloes. The practice of using 
halo merger histories to understand the build up of galaxies can be traced back to 
\citet{bower91}, \citet{cole91}, and \citet{kauffmannWG93}. In these pioneering works 
the growth of haloes is described by analytical methods. Numerical techniques like 
N-body numerical simulations can deal more accurately with the gravitational 
processes underlying the evolution of cosmic structure. The clustering of haloes is tracked, 
snapshot by snapshot, and stored in a tree form (`merger tree'). Halo merger trees therefore 
record, in a direct way, when and how haloes assemble by accreting other building blocks, 
and are widely used to rebuild galaxy assembly histories \citep[e.g.][]{kauffmannWG93, 
roukemaEtal97, kauffmannEtal99, springelEtal01}. 

To compute galaxy merger rates, one possibility is to combine the halo merger trees with 
a redshift-dependent abundance matching model that statistically assigns galaxies to dark matter haloes 
\citep{fakhouriM08, mosterNW13, behrooziWC13}. In this fashion, the observed abundance of 
galaxies can be inverted to estimate the galaxy merger rate as a function of halo mass and 
redshift. This provides a great deal of insight, but relies on the accuracy of the statistical 
model. Although appealing because of its close relation to the real data, the approach may 
miss physical correlations between the merging objects. A preferable approach is therefore to 
form galaxies within dark matter haloes using a physical  galaxy formation model. It is important to note, 
however, that reliable conclusions can only be obtained if the overall galaxy stellar mass function 
accurately reproduces observational measurements \citep{bensonEtal03, schayeEtal15}. 

One approach is to use `semi-analytic' models of galaxy formation. By introducing 
phenomenological descriptions for feedback from star formation and AGN, such models are 
able to reproduce the observed galaxy stellar mass function (e.g. \citealt{bowerEtal06,crotonEtal06}, 
for a recent review see \citealt{knebeEtal15}). \cite{deluciaEtal06} study the assembly of elliptical 
galaxies in a semi-analytic model based on the model of \cite{crotonEtal06}. They find 
that stars in massive galaxies (with stellar mass $M_*\geq10^{11}\msun$) are formed earlier ($z\gtrsim2.5$) but 
are assembled later (by $z\approx0.8$). \cite{deluciaB07} show further that massive 
members in galaxy clusters assemble through mergers late in the history of the Universe, 
with half of their present-day mass being in place in their main progenitor by $z\approx0.5$. 
In contrast, less massive galaxies undergo relatively few mergers, acquiring only $20\%$ of their 
final stellar mass from external objects. \cite{parryEF09} study 
the assembly and morphology of galaxies in the semi-analytic model of \cite{bowerEtal06}. 
They found many similarities, but also important disagreements, stemming primarily from 
the differing importance of disk instabilities in the two models. \cite{parryEF09} find that major 
mergers are not the primary mass contributors to most spheroids except the brightest ellipticals. 
This instead is brought in by minor mergers and disk instabilities. In their model, the majority 
of ellipticals, and the overwhelming majority of spirals, never experience a major merger. 

Semi-analytic studies such as those above give important insights but suffer from the limitations 
inherent to the approach, for example, the neglect of tidal stripping of infalling satellites and 
the absence of information about the spatial distribution of stars, as well as being limited 
by the overall accuracy of the model. Numerical simulations have fewer limitations, 
and have thus become an alternative useful tool for these studies. \cite{hopkinsEtal10} 
compare the galaxy merger rates derived from a variety of analytical models and hydrodynamical 
simulations. They find that the predicted galaxy merger rates depend strongly on the prescriptions for 
baryonic physical processes, especially those in satellite galaxies. For example, the lack of 
strong feedback can result in a difference in predicted merger rates by as much as a factor of 5. 
Mass ratios used in merger classification also have an impact on merger rate prediction. 
Using the stellar mass ratio, rather than the halo mass ratio, can result in an order of magnitude change in the derived merger rate. 

With rapidly increasing computational power and much progresses in modelling physical processes 
on sub-grid scales, cosmological N-body hydrodynamical simulations are increasingly capable of 
capturing the physics of galaxy formation \citep[e.g.][]{hopkinsEtal13,vogelsbergerEtal14}. 
The Evolution and Assembly of Galaxies and their Environments (EAGLE) simulation project accurately 
reproduces the observed properties of galaxies, including their stellar mass, sizes and formation
histories, within a large and representative cosmological volume \citep{schayeEtal15,furlongEtal15b, 
furlongEtal15}. This degree of fidelity makes the EAGLE simulations a powerful tool for 
understanding and interpreting a wide range of observational measurements. 
Previous papers have focused on the evolution 
of the mass function and the size distribution of galaxies \citep{furlongEtal15b,furlongEtal15}, 
the luminosity function and colour diagram \citep{trayfordEtal15} and galaxy rotation curves 
\citep{schallerEtal15b} as well as many aspects of the HI and H$_2$ distribution of 
galaxies \citep{lagosEtal15, baheEtal15, crainEtal16} in the EAGLE universe. But none has tracked the assembly of 
individual galaxies and decipher the underlying mechanisms as yet. 
As an attempt to shed some light on the issue, in this work we connect galaxies seen at different redshifts, 
creating a merger tree that enables us to 
establish which high-redshift fragments collapse to form which present-day galaxies (and vice versa). 
In this way we can quantify the importance of in-situ star formation relative to the mass gain 
from galaxy mergers and diffuse accretion. Throughout the paper, we will focus on the main, 
or `central', galaxies, avoiding the complications of 
environmental processes such as ram pressure stripping and strangulation that suppress star formation and 
strip stellar mass from satellites. Unless otherwise stated, stellar masses refer to the stellar mass of 
a galaxy at the redshift of observation, not to the initial mass of stars formed. 

The outline of this paper is as follows. In Section~\ref{sim}, we provide a brief overview of the numerical 
techniques and subgrid physical models employed by the EAGLE simulations, and describe the 
methodology used to construct merger trees from simulation outputs. We investigate the assembly 
histories and merger histories of galaxies and discuss the impact of feedback on galaxy mass build-up in Section~\ref{res}. 
We compare our results with some previous works in Section~\ref{dis}, and finally 
summarize in Section~\ref{con}. The appendices present the detailed criteria we use 
to define galaxy mergers and show the impacts of our choices of galaxy mass on our results. 
The cosmological parameters used in this work is 
from the {\it Planck} mission \citep{PlanckEtal14} , 
$\Omega_{\Lambda}=0.693$, $\Omega_m=0.307$, $h=0.677$, $n_s=0.96$ and $\sigma_8=0.829$. 
  
\section{EAGLE simulation and merger tree}\label{sim} 
\subsection{EAGLE simulation}\label{proj} 
The galaxy samples for this study are selected from the Evolution and Assembly of Galaxies and 
their Environments (EAGLE) simulation suite \citep{schayeEtal15,crainEtal15}. The EAGLE simulations 
follow the evolution (and, where appropriate, the formation) of dark matter, gas, stars and black holes 
from redshift $z=127$ to the present day at $z=0$. They were carried out with a modified version of 
the GADGET 3 code \citep{springel05} using a pressure-entropy based formulation of 
Smoothed Particle Hydrodynamics method \citep{Hopkins13}, coupled to several other improvements to 
the hydrodynamic calculation (Dalla Vecchia (in preparation); \citealt{schayeEtal15,schallerEtal15a}). 
The simulations include subgrid descriptions for radiative cooling \citep{WiersmaSS09}, 
star formation \citep{SchayeDV08}, multi-element metal enrichment \citep{WiersmaEtal09}, 
black hole formation \citep{SpringelDMH05, rosasguevaraEtal15} as well as feedback from 
massive stars \citep{dallavecchiaS12} and active galactic nuclei (AGN) \citep[for a complete description, see][]{schayeEtal15}. 
The subgrid models are calibrated using a well-defined set of local observational 
constraints on the present-day galaxy stellar mass function and 
galaxy sizes \citep{crainEtal15}. 

Each simulation outputs 29 snapshots to store particle properties over the redshift range $0\leq z\leq20$. 
The corresponding time interval between snapshot outputs ranges from ${\sim}0.3$ to ${\sim}1.35$~Gyr. 
The largest EAGLE simulation, hereafter referred to as Ref-L100N1504, employs 
$1504^3$ dark matter particles and an initially equal number of gas particles in a periodic cube with 
side-length $100$~comoving Mpc (cMpc) on each side. This setup results in a particle mass of 
$9.7\times 10^6\msun$ and $1.81\times 10^6\msun$ (initial mass) for dark matter and gas particles 
respectively. The gravitational force between particles is calculated using a Plummer potential with 
a softening length set to the smaller of $2.66$~comoving kpc (ckpc) and 0.7~physical kpc (pkpc). 

The formation of galaxies involves physical processes operating on a huge range of scales, 
from the gravitational forces that drive the 
formation of large scale structure on $10-100$ Mpc scales, to the processes that lead to the formation of 
individual stars and black holes on $0.1$ pc and smaller scales. Such a dynamic range, $10^9$ in length and 
perhaps $10^{27}$ in mass, cannot be computed efficiently without the use of subgrid models.
Such models are inevitably approximate and uncertain. In EAGLE, we require that the subgrid models are 
physically plausible, numerically stable and as simple as possible. The uncertainty in these models introduces 
parameters whose values must be calibrated by comparison to observational data \citep{VernonGB10}. 
We explicitly recognise that these models are approximate and adopt the clear methodology for 
selecting parameters and validating the model that is described in detail in \cite{schayeEtal15} and \cite{crainEtal15}.  
The subgrid parameters calibrated by requiring that the model fits three key properties of local galaxies well: 
the galaxy stellar mass function, the galaxy size -- mass relation and the normalisation of the black hole mass -- 
galaxy mass relation and that variations of the parameters alter the simulation outcome in predictable ways \citep{crainEtal15}. 
We find that these data-sets can be described well with physically plausible values for the subgrid parameters. 
We then compare the simulation with further observational data to validate the simulation. We find that 
it describes many aspects of the observed universe well (i.e., within the plausible observational uncertainties), 
including the evolution of the galaxy stellar mass function and 
star formation rates \citep{furlongEtal15}, evolution of galaxy colours and luminosity functions \citep{trayfordEtal15}. 
It also provides a good match to observed OVI column densities \citep{rahmatiEtal16} and 
molecular content of galaxies \citep{lagosEtal15} , as well as a reasonable description of the X-ray luminosities of 
AGN \citep{rosasguevaraEtal15}. The good agreement with these diverse datasets, especially those distantly 
related to the calibration data, provides good reason to believe that the simulation provides 
a good description of the evolution of galaxies in the observed Universe. 
It can therefore be used to explore galaxy assembly histories in ways that are not accessible
to observational studies.

\subsection{Halo identification and subhalo merger tree}\label{halotree}
Building subhalo merger trees from cosmological simulations involves two steps: Firstly, we identify 
haloes and subhaloes as gravitationally self-bound structures; Secondly, we identify 
the descendants of each subhalo across snapshot outputs and establish the 
descendant-progenitor relationship over time.

\subsubsection{Halo identification}\label{halo}
Dark matter structures in the EAGLE simulations are initially identified using the  
\lq\lq{}Friends-of-Friends\rq\rq{} (FoF) algorithm with a linking length of 0.2 
times the mean inter-particle spacing \citep{davisEtal85}. Other particles 
(gas, stars and black holes) are assigned to the same FoF group as 
their nearest linked dark matter neighbours. The gravitationally bound substructures 
within the FoF groups are then identified by the SUBFIND algorithm \citep{springelEtal01,dolagEtal09}.
Unlike the FoF group finder,  SUBFIND considers all species of particle and identifies self-bound 
subunits within a bound structure which we refer to as `subhaloes'. Briefly, the algorithm assigns 
a mass density at the position of every particle through a kernel interpolation over a certain number of 
its nearest neighbours. The local minima in the gravitational potential field are the centres of subhalo candidates. 
The particle membership of the subhaloes is determined by the iso-density contours defined by 
the density saddle points. Particles are assigned to at most one subhalo. The subhalo with a minimum value of 
the gravitational potential within a FoF group is defined as the main subhalo of the group. 
Any particle bound to the group but not assigned to any other subhaloes within the group are assigned to the main subhalo. 
    
\subsubsection{Subhalo merger tree}\label{tree}
Although they orbit within a FoF group, subhaloes survive as distinct objects for an extended period of time. 
We therefore use subhaloes as the base units of our merger trees: FoF group merger trees can be rebuilt 
from subhalo merger trees if required. 
The first, and main, step in building the merger tree is to link subhaloes across snapshots. 
As in \cite{springelEtal05}, we search the descendant of a subhalo by tracing the most 
bound particles of the subhalo. We use the D-Trees algorithm \citep{jiangEtal14} to locate 
the whereabouts of the $N_{link}=min(N_{linkmax}, max(f_{trace}N, N_{linkmin}))$ 
most bound particles of the subhalo, where $N$ is the total particle number in the subhalo. 
We use parameters $N_{linkmin}=10$, $N_{linkmax}=100$, $f_{trace}=0.1$ in the descendant search. 
The advantages of focusing on the $N_{link}$ most bound particles are twofold. On the one hand, 
D-Trees can identify a descendant even if most particles are stripped away leaving only 
a dense core. On the other hand, the criterion minimises misprediction of mergers during 
flyby encounters \citep{fakhouriM08,genelEtal09}.
 
The descendant identification proceeds as follows. For a subhalo $A$ at a given snapshot, 
any subhalo at the subsequent snapshot that receives at least one particle from $A$ is labelled 
as a descendant candidate. From those candidates we pick the one that receives the 
largest fraction of $A$\rq{}s $N_{link}$ most bound particles (denoted as $B$) as the 
descendant of $A$. $A$ is the progenitor of $B$. If $B$ receives a larger fraction of 
its own $N_{link}$ most bound particles from $A$ than from any other subhalo at previous snapshot, 
$A$ is the principal progenitor of $B$. A descendant can have more than one progenitor, 
but only one principal progenitor. The principal progenitor can be thought of
as `surviving' the merger while the other progenitors lose their individual identity. 

Subhaloes sometimes exhibit unstable behaviour during mergers, 
complicating the descendant/progenitor search. When a subhalo passes through the dense core of 
another subhalo, it may not be identifiable as a separate object at the next snapshot, 
but will then reappear in a later snapshot. From a single snapshot, there is no way to know whether 
the subhalo has merged with another subhalo, or has just disappeared temporarily, and we need to 
search a few snapshots ahead in order to know which case it falls into. In practice we search up to $N_{step}=5$ 
consecutive snapshots ahead for the missing descendants. This gives us between one and $N_{step}$ 
descendant candidates. If the subhalo is the principal progenitor of one or more candidates, the earliest candidate 
that does not have a principal progenitor is chosen to be the descendant. If there is no such candidate, 
then the earliest one will be chosen. If the subhalo is not the principal progenitor of 
any candidates, it will be considered to have merged with another subhalo and no longer appears 
as an identifiable object.

\begin{figure}  
\centering\includegraphics[width=8cm,angle=0]{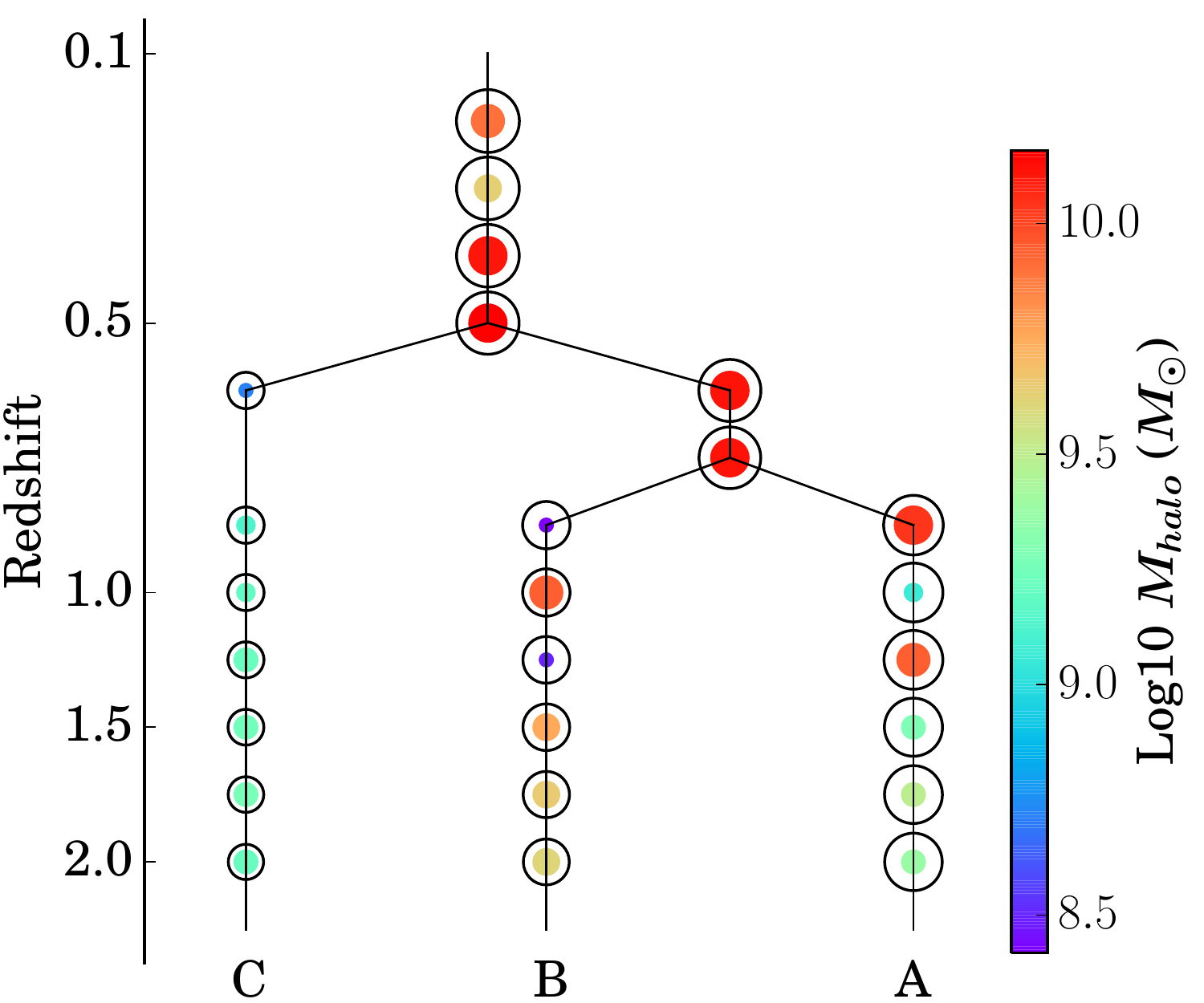}\vspace{-0.1cm}
\caption{A section of a subhalo merger tree illustrating how subhaloes following branches A and B 
exchange particles before merging. The colour of the solid symbol reflects the halo mass, while 
the size of the circle represents the `branch mass', which is the sum of the total mass of 
all the progenitors sitting on the same branch. A see-saw behaviour is clearly seen in the evolution of the 
halo mass, which may confuse identification of the most important branch. 
Instead we use branch mass to locate the main branch of the tree. In this plot, Branch A has the largest 
branch mass and is therefore chosen as the main branch, even though its progenitors 
are not always the most massive ones.} \label{fig_swap}
\end{figure}

Occasionally two subhaloes enter into a competition for bound particles. This occurs as the
participants orbit each other prior to merging. In SUBFIND the influence of a subhalo is based on 
its gravitational potential well. When two subhaloes are 
close to each other, their volumes of influence become intertwined and the definition of the
main halo may become unclear. For example, when a satellite subhalo
orbits closely to its primary host, the satellite can be tidally compressed at some stage and become 
denser than the host. At this point, the satellite may be classified as the central object of the halo so
that most of the halo particles are assigned to it. At a later time, the original central, however, 
can surpass the satellite in density and reclaim the halo particles. This contest can last for several 
successive snapshots, accompanied by a see-saw exchange of their physical properties 
during the merging. Fig.~\ref{fig_swap} shows an example in which merging haloes take 
turns to be classified as the central host during the merging process. Overall, 
fewer than $5\%$ of subhalo mergers in the EAGLE simulations exhibit this behaviour, 
compatible with the statistics found by \cite{wetzel09}. 
The fact that a fierce contest between subhaloes is sometimes seen during the merging process 
highlights the inherent difficulties in appropriately describing subhalo properties at that stage. 

The property exchanges during such periods are not physical, but rather stem from the 
requirement that particles be assigned to a unique subhalo on the basis of the spatial coordinates 
and the local density field in a single snapshot. The history of an object is, however, conveniently 
simplified by modifying the definition of the most massive progenitor to account for its mass in earlier snapshots. 
We refer to this progenitor as the `main progenitor', and the branch they stay on in the object\rq{}s merger tree 
as the `main branch'. Because of the mass exchange discussed above, we track the main branch using 
the `branch mass',  the sum of the mass over all particle species of all progenitors on the same branch 
\citep{deluciaB07}. The main progenitor is then the progenitor that has the maximum 
branch mass among its contemporaries. This can avoid the misidentification of main progenitors due to 
the property exchanges occurring for merging subhaloes as we see in Fig.~\ref{fig_swap}. 
It is worth noting that according to this definition, a lower-mass progenitor which has existed for a 
long time can sometimes be preferred over a more massive progenitor which has formed quickly, 
when locating main progenitors.  

The subhalo merger trees derived by the method described above are publicly 
available through an SQL database \footnote{http://www.eaglesim.org} 
similar to that used for the Millennium simulations \citep[see][for more details]{mcalpineEtal15}.

\subsection{Galaxy sample, galaxy merger tree, and merger type}\label{galm}
\begin{figure*} 
\centering\includegraphics[width=16cm,angle=0]{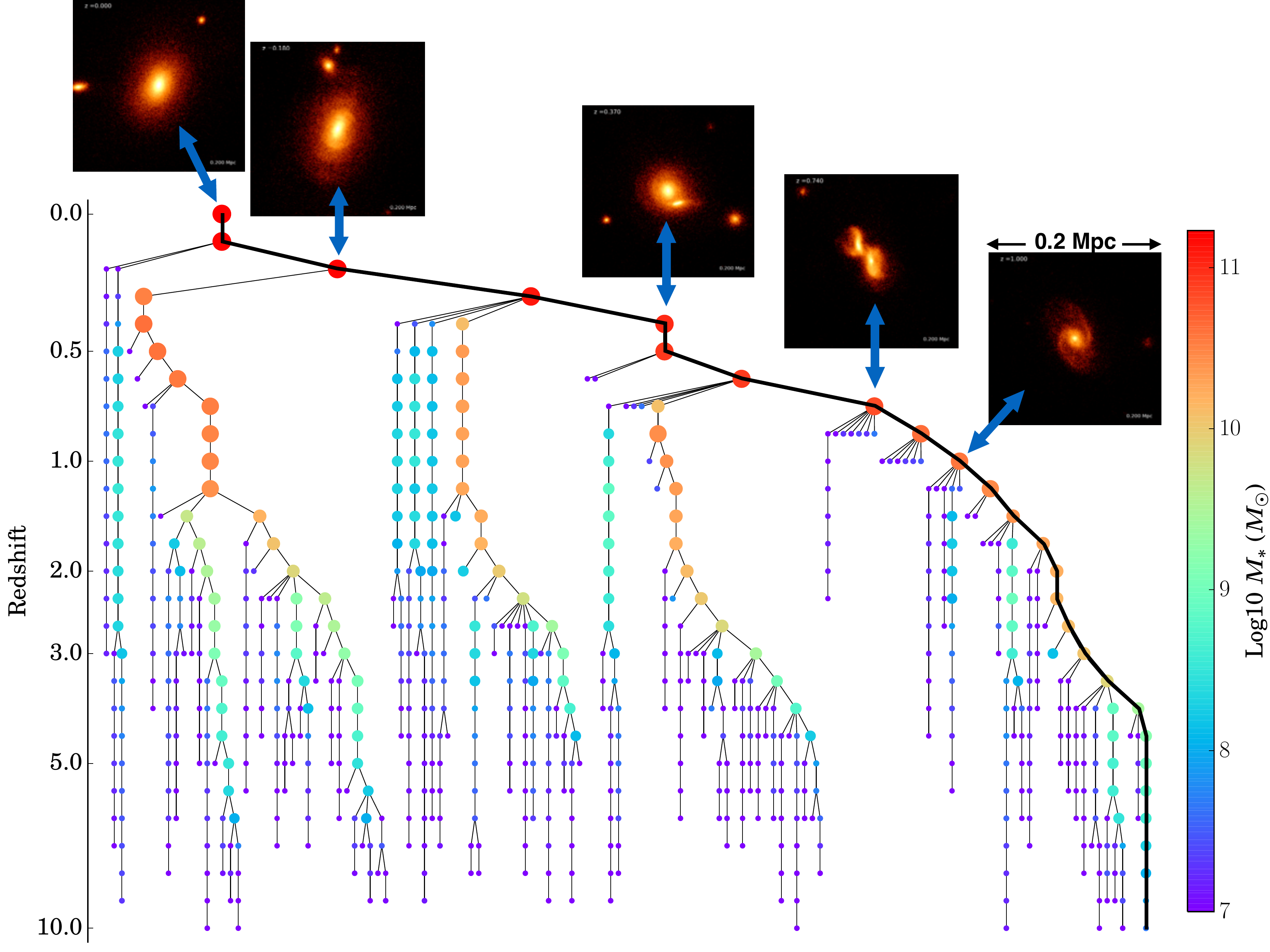}\vspace{-0.1cm}
\caption{An example of a galaxy merger history. The galaxy has a stellar mass $M_*=1.7\times10^{11}\msun$ 
 at redshift $z=0$. Symbol colours and sizes are logarithmically scaled with stellar mass. The thick solid line marks 
 the main branch. The final galaxy is built from many small progenitors, but most of these contributors have very low mass. 
 We also show images of its stellar mass distribution in a $200$~comoving kpc box at a few redshifts. The
galaxy shows prominent spiral-like structure at redshift $z=1$, but then experiences several interactions with 
other objects, passing through a shell-like phase to transform into an elliptical galaxy at $z=0$.}\label{fig_tree}
\end{figure*}

In this work galaxies are identified as the stellar components of the subhaloes. The main subhalo of a FoF halo 
hosts the `central' galaxy, while other subhaloes within the group host satellite galaxies. We will focus on the central galaxies 
in our study, avoiding the complications of environmental processes such as ram pressure stripping and strangulation 
that suppress star formation and strip stellar mass from satellite galaxies 
\citep[e.g.][]{wetzelEtal13,McGeeEtal14, BarberEtal16}. 

The stellar mass of a galaxy is measured using a spherical aperture. This gives similar results to 
the commonly-used 2-D Petrosian aperture used in observational work, but provides an 
orientation-independent mass measurement for each galaxy. Previous studies based on the 
EAGLE simulations adopt an aperture of $30$~pkpc to measure galaxy stellar mass 
\citep[e.g.][]{schayeEtal15, furlongEtal15}. Nevertheless, subhaloes do contain a significant 
population of diffuse stars, particularly in more massive haloes \citep{furlongEtal15}. Such 
stars are probably deposited by interactions and tidal stripping, and sometimes observed as 
low surface brightness intracluster/intragroup light \citep{theunsW97, zibettiW04,mcgeeB10}. 
Since the formation of massive galaxies is a particular focus of this paper, we use a larger aperture, 
with a radius of $100$~pkpc, to calculate galaxy mass. Note that this mass does not include 
the stellar mass of satellites lying within the $100$~pkpc aperture. As we will show in Appendix~\ref{ap_ap}, 
this aperture choice has little impact on galaxy properties for galaxies with stellar mass 
$M_*<10^{11}\msun$ \citep[see also][]{schayeEtal15}.  

Unless otherwise stated, the galaxy stellar mass in this work refers to the actual mass of stars 
in the galaxy at the epoch of  `observation'. Using actual mass replicates what an ideal observer would
measure and directly addresses the question of when the current stellar population of the galaxy was
formed/assembled. Nevertheless, we should note that the mass budget of the current stellar population 
is a combination of two processes:  stellar mass gain via star formation, 
accretion and merging, and mass loss through stellar evolution processes. However, using the actual
stellar mass complicates interpretation of the relative mass contribution from different types of merger events
since it depends on the age of the stellar population that is accreted. We therefore use the stellar mass
initially formed (`initial mass'), not the actual stellar mass, to evaluate the 
contributions from internal and external processes to galaxy assembly. In practice, this distinction has little
effect on the results and we show the effect of using initial stellar mass throughout in Appendix~\ref{ap_tfta}.

\subsubsection{Galaxy sample}
Our study is based on the formation histories of 62,543 galaxies in the largest EAGLE simulation Ref-L100N1504, 
spanning a stellar mass range of $10^{9.5}-10^{12}\msun$ over redshift $z=0-3$. In order to test the robustness of 
our results to resolution, we also extract 1,381 galaxies within the same mass range, as a comparison sample, 
from the EAGLE simulation Recal-L025N0752 ($2 \times 752^3$ dark matter and gas particles in a $25$ cMpc box), 
which has 8 times better mass resolution and the same snapshot frequency as Ref-L100N1504. 
We use subgrid physical models with parameters recalibrated to the present-day observations, 
as this provides the best match to the observed galaxy population \citep[see][]{schayeEtal15}. 
In order to study the mass dependence of galaxy assembly, we split our samples into three stellar mass 
bins: a low-mass bin ($10^{9.5}\leq M_*<10^{10.5}\msun$), an intermediate-mass bin ($10^{10.5}\leq M_*<10^{11}\msun$) 
and a high-mass bin ($10^{11}\leq M_*<10^{12}\msun$) . 

\subsubsection{Galaxy merger tree}
We construct galaxy merger trees by focusing on the stellar component of the subhalo merger trees. 
Fig.~\ref{fig_tree} shows such a tree for a galaxy with $M_*=1.7\times10^{11}\msun$ at $z=0$, together 
with images of its star  distribution highlighting its morphological evolution since $z=1$. The main branch 
of the tree is marked by the thick black line. It is important to bear in mind that the identification of the 
main branch is always based on the branch mass; at any particular epoch, the most massive galaxy progenitor may not lie on 
the main branch. However, for the reasons described in Section~\ref{tree}, using the branch mass yields more stable and 
intuitive results. 

Galaxy merger trees appear broadly similar to subhalo merger trees, except that  
the latter contain more fine branches corresponding to small subhaloes within which no stars 
have formed. Galaxy trees are also less effected by the mass exchange issue than subhalo trees, 
as star particles are more spatially concentrated.

\subsubsection{Merger type}\label{merger}
The effects of tidal forces and torques during a merger depend on the mass ratio of the merging systems 
\citep[e.g.][]{barnesH92}. A merger between a low-mass satellite and a more massive 
host is generally less violent than a merger between systems of comparable mass, and has a less dramatic impact 
on the dynamics and morphology of the host. It is therefore useful to classify mergers 
into different types according to the mass ratio between the two merging systems, 
$\mu\equiv M_2/M_1$ ($M_1>M_2$). For galaxy mergers, $\mu$ is the ratio of stellar masses between two merging 
galaxies. While for halo mergers, it is the halo mass ratio. 

While this is straightforward in semi-analytic models 
(since galaxies are uniquely defined entities), in numerical simulations (and in nature as well) merging 
systems experience mass loss due to tidal stripping throughout the merging process. Our strategy is therefore 
to choose a separation criterion, $R_{merge}$, and determine the merger type when the 
merging systems are separated, for the first time, by that distance or less. For galaxy mergers, 
we adopt $R_{merge}=5\times R_{1/2}$ where $R_{1/2}$ is the half-stellar mass radius of 
the primary galaxy (note that $R_{merge}$ is not a projected but a 3D separation). 
The value of $R_{merge}$ ranges from $\sim20$ to $200$~pkpc in the stellar mass range 
explored in this work (see Appendix~\ref{ap_r}), and is similar to the projected separation criteria 
adopted in observational galaxy pair studies. For subhalo mergers, $R_{merge}=r_{200}$, where 
$r_{200}$ is the radius of a region around the FoF group of the subhaloes within which the density is 
200 times the cosmological critical density. In the rare event that an object is located within the 
$R_{\rm merge}$ of more than one other object, it is considered to be the merging companion of the 
nearest one. 

More often than not, the secondary object may have suffered tidal stripping of mass 
when the merger type is determined due to the finite time sampling of our snapshot outputs. 
To alleviate the resulting misestimate of the mass ratio, we compare the mass of the merging systems 
at the start of the merging event with that at the previous snapshot, and use the maximum 
to calculate the mass ratio $\mu$. In our study, merging events are classified as {\it major mergers} if 
$\mu\geq 1/4$; as {\it minor mergers} if $1/4>\mu\geq 1/10$; and as diffuse {\it accretion}, when 
$\mu<1/10$. Our major merger definition is different from that of \cite{coleEtal00} or \cite{deluciaB07} 
who adopt a larger mass ratio $\geq 1/3$, but is similar to more recent studies \cite[e.g.][]{rodriguezgomezEtal15}. 
Mergers with mass ratio $\geq1/4$ can produce strong asymmetries in the morphology of both merging galaxies, 
making them easily identifiable in observations \citep{casteelsEtal14}. 

\begin{figure}
\centering\includegraphics[width=8.5cm,angle=0]{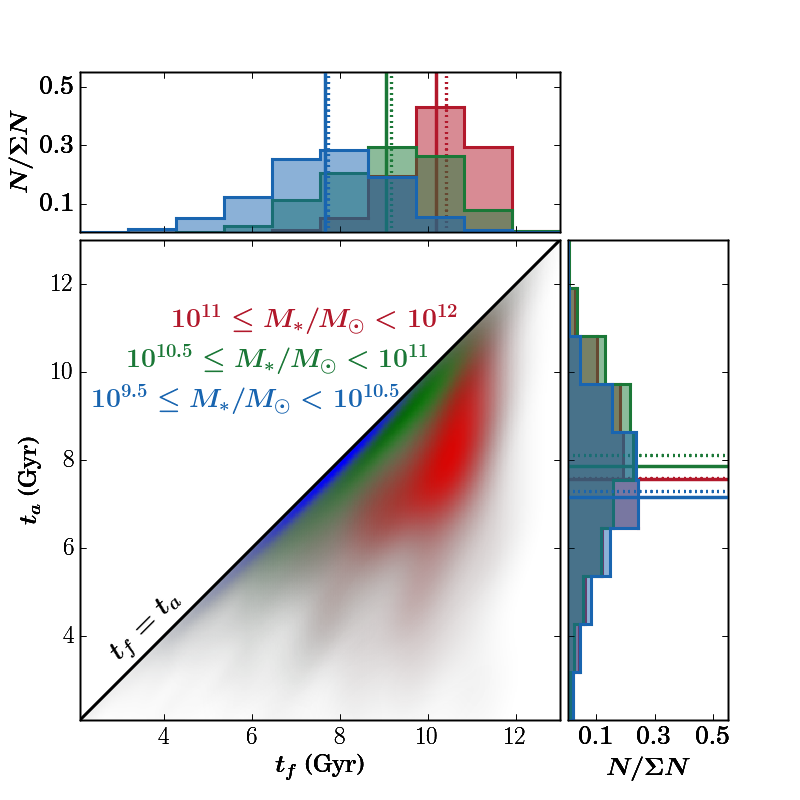}\vspace{-0.cm}
\caption{The formation time, $t_f$, as a function of assembly time, $t_a$, for galaxies at $z=0$. Both timescales are 
measured as lookback times in Gyr, and the galaxies are classified into three bins of stellar mass by colours. The solid line 
represents the one to one relation for the two timescales.  Galaxies with stellar mass ($M_*\leq10^{10.5}\msun$), are 
distributed along this line, indicating that they assemble most of their stars through in-situ star formation. 
In massive galaxies of $M_*>10^{11}\msun$, by contrast,
$t_a$ lags behind $t_f$, and the galaxies are offset from the $t_a=t_f$ line, showing the importance of stars formed in other 
objects and subsequently accreted. The normalised histograms of the $t_f$ and $t_a$ distributions are shown in marginal panels. The mean and the median of the distributions are indicated by the solid and dotted lines, respectively.}\label{fig_tfta}
\end{figure}

\begin{figure}
\centering\includegraphics[width=8.5cm,angle=0]{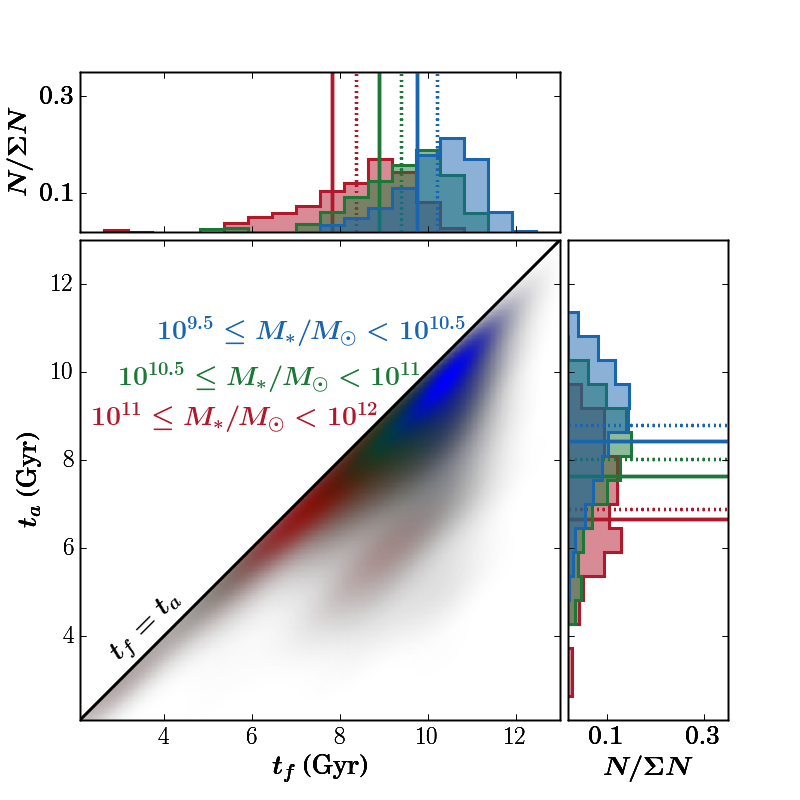}\vspace{-0.cm}
\caption{Formation and assembly times for the parent dark matter halo of the galaxies shown in Fig~\ref{fig_tfta}. 
Note that haloes are binned by the stellar mass of their central galaxy, but that bins of higher stellar mass correspond to 
higher mean halo mass. The solid line represents the case where $t_f=t_a$, as in Fig~\ref{fig_tfta}. 
In contrast to the situation for galaxies, $t_f$ and $t_a$ increase with decreasing stellar mass, demonstrating 
the hierarchical nature of the mass assembly of dark matter haloes. Note, however, that formation and 
assembly times are similar regardless of mass, meaning that halo growth is dominated by accretion of diffuse material. 
The assembly histories of haloes are markedly different from those of the galaxies they contain.}\label{fig_h_tfta}
\end{figure}

\section{Results} \label{res}
\subsection{Galaxy formation and assembly timescales}\label{tfta}
A simple way to summarise the formation history of a galaxy is to measure the timescale on which 
it assembles its mass. As discussed by \cite{deluciaEtal06} and \cite{neisteinBD06}, this can be assessed in 
two ways. Firstly we can measure the total stellar mass in all progenitors of the final galaxy as a function of time. 
This mass increases through star formation. For many purposes, however, it is more relevant to focus on 
the growth of the main progenitor if we are interested in connecting galaxies identified in observational studies 
at different epochs. Following \cite{deluciaEtal06}, we refer to the timescale by which the total mass of all progenitors 
has reached half of the stellar mass of the final galaxy as 
the `formation time', $t_f$. $t_f$ is closely related to the star formation history of the galaxy. 
The timescale by which the main progenitor of the final galaxy has assembled that much mass is defined 
as the `assembly time', $t_a$. Both timescales are measured in lookback times. If a galaxy forms 
most of its stars through in-situ star formation, it will have $t_f\approx t_a$, 

Fig.~\ref{fig_tfta} plots formation time, $t_f$, against assembly time, $t_a$, for galaxies at $z=0$ in three 
stellar mass bins. The galaxies occupy different regions in the plot depending on their stellar mass. 
Low-mass galaxies ($M_*<10^{10.5}\msun$) typically formed their stars 8~Gyr ago. In spite of a large spread, 
their formation times scatter about the line of $t_a=t_f$, implying an in-situ origin for their stars. 
In contrast, the most massive galaxies formed their stars relatively early,  
$t_f\sim11$~Gyr, and have $t_a<t_f$ indicating that a fraction of their stars are formed 
elsewhere and subsequently assembled into the final system. The delay between $t_a$ and $t_f$ 
is a strong function of galaxy mass, increasing rapidly as the galaxy mass exceeds $10^{11}\msun$.   
This trend agrees well with previous work \citep[e.g.][]{deluciaEtal06,neisteinBD06,parryEF09}.
It is also seen in observational data, as a trend referred to as `downsizing', where old stellar populations 
dominate massive galaxies \citep{bowerLE92,cowieEtal96,heavensEtal04,gallazziEtal05} and 
low-mass galaxies have a more extended period of star formation \citep{noeskeEtal07a,leitner12}. 
These results hint that most low-mass galaxies that formed at high redshifts do not `survive' to the 
present day and have merged into more massive galaxies. Indeed, we find that only half of the galaxies with 
$M_*\sim 10^{9}-10^{10.5}\msun$ at $z=3$ survive to $z=0$. 

In a $\Lambda$CDM universe, dark matter haloes grow in a self-similar manner, 
with high-mass haloes typically being formed more recently than their low-mass counterparts \citep{davisEtal85, bardeenEtal86}. This is confirmed in Fig.~\ref{fig_h_tfta}, which shows the distribution of $t_a$ as a function of 
$t_f$ for the haloes hosting the galaxies of Fig.~\ref{fig_tfta}. Points are again 
coloured by the present-day stellar mass of the galaxies, as in Fig.~\ref{fig_tfta}. 
We see that both timescales decrease with increasing halo mass, as expected from the 
hierarchical structure formation scenario. This is entirely the opposite trend to that seen for the galaxies. 

This apparent contradiction is the result of AGN feedback being more effective in high-mass 
haloes \citep{bowerEtal06}. At low mass, stellar feedback causes the galaxy\rq{} stellar 
mass to scale with approximately the square of the halo mass, so that the galaxies grow rapidly as 
the halo mass increases. The stars gained by accretion and merging are dwarfed by the contribution 
from ongoing star formation. 
However, once the halo mass exceeds $\sim 10^{11.5}\msun$ star formation is strongly suppressed 
by AGN feedback (see \citealt{rosasguevaraEtal15} and Bower et al. (in preparation))
and galaxies grow almost exclusively by accretion and mergers. 
This transition breaks any self-similarity in the hierarchy: although the most massive galaxies 
assemble late, the stars they contain were formed at much earlier epochs.

The halo assembly and formation times are remarkably close. This occurs because the dominant 
contribution to halo growth comes from matter which is not yet bound into galaxy-bearing dark matter 
haloes. Many previous studies have pointed out that in a CDM cosmology halo growth is driven by a mix of mergers and 
accretion of matter that has not yet collapsed into identifiable haloes 
\citep[e.g.][]{laceyC93,kauffmannW93,guoW08, fakhouriM10,genelEtal10, wangEtal11}. In contrast 
stars are only formed efficiently in well defined massive haloes. This fundamental differences results in the stark contrast between  Fig.~\ref{fig_tfta}  and Fig.~\ref{fig_h_tfta}.

\subsection{The redshift evolution of galaxy formation and assembly times}\label{dt}
In previous section we have shown that the delay between formation time and assembly time 
can provide some useful hints on how a galaxy assemble its mass. In this section, 
we use the differences of both timescales as a tool to examine the assembly history of 
galaxies at different redshifts. 

\begin{figure}  
\centering\includegraphics[width=8cm,angle=0]{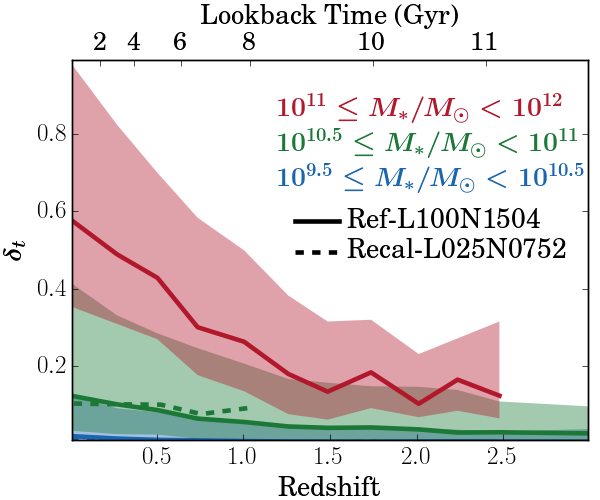}\vspace{-0.cm}
\caption{The evolution of the relative difference between the assembly and formation times, $\delta_t$,  
for galaxies in three stellar mass bins (indicated by colour and legend). Lines represent the medians of the 
$\delta_t$ distributions. The shaded regions enclose the 25th to 75th percentiles. Bins with fewer than 10 
galaxies are not shown. $\delta_t$ increases with stellar mass but decreases with redshift, showing the 
importance of external processes in the mass assembly of low-redshift massive galaxies. These trends are 
insensitive to resolution, as shown by the agreement between the results of Ref-L100N1504 ({\it solid lines}) 
and of Recal-L025N0752 ({\it dashed lines}), although the latter simulation lacks objects in the highest-mass bin.}\label{fig_dtmz}
\end{figure}

To quantity the relative difference between the two timescales, we define a dimensionless parameter, 
\[\delta_t\equiv 1-t_f/t_a\]
Fig.~\ref{fig_dtmz} shows the distribution of $\delta_t$ for galaxies at redshift $z=0-3$. 
As before these galaxies are split into three stellar mass bins. We show results for 
Ref-L100N1504 (solid lines), as well as for the higher-resolution (but smaller volume) simulation 
Recal-L025N0752 (dashed lines) in order to demonstrate the convergence of the results. 
The shaded region represents the 25th to 75th percentiles of the $\delta_t$ distribution. 
While low-mass galaxies have median $\delta_t<0.1$ at all redshifts, high-mass galaxies have 
median $\delta_t$ decreases with increasing redshift, showing that stellar accretion loses 
ground to in-situ star formation. The same redshift dependence is also found in 
semi-analytic studies \citep[e.g.][]{guoW08}. This evolution results from the 
much higher specific star formation rates of high-redshift galaxies due (at least in part) to the 
higher gas infall rates and the less efficient AGN feedback of young galaxies.

\begin{figure*}  
\centering\includegraphics[width=16cm,angle=0]{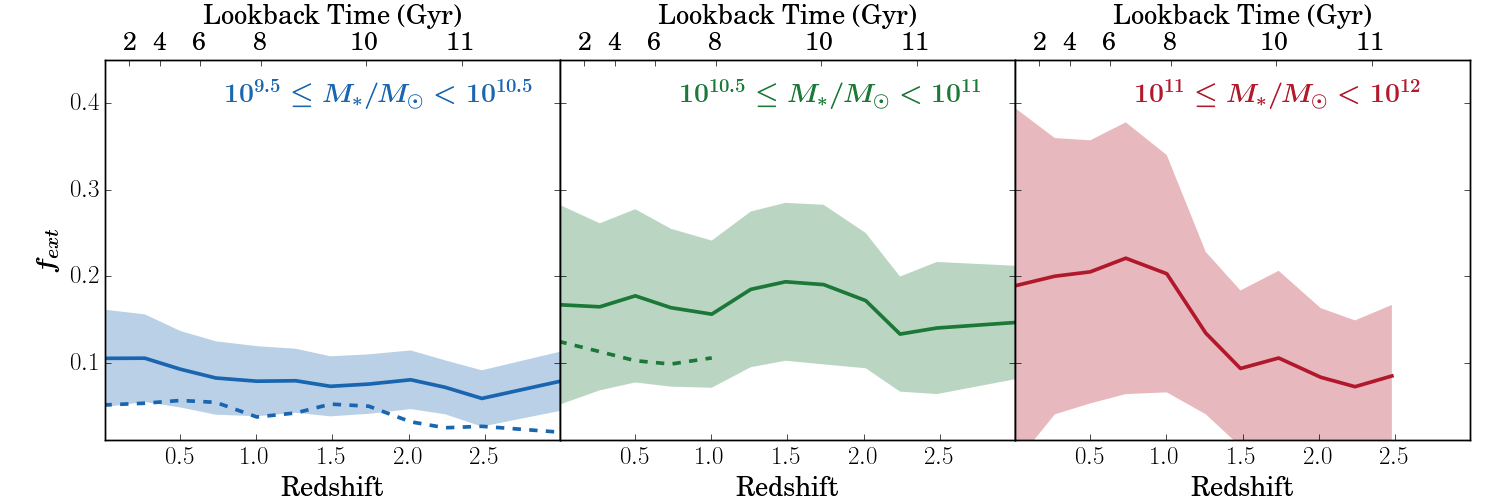}\vspace{-0.cm}
\caption{The initial stellar mass contributions of mergers (i.e. mass ratio $\mu\geq 1/10$) and diffuse accretion ($\mu< 1/10$) for galaxies of different stellar mass, 
at redshifts $z=0-3$ in three stellar mass bins (as coloured). Solid and dashed lines represent the median of the distribution 
in Ref-L100N1504 and Recal-L025N0752, respectively. Our analysis stops at the redshift when fewer than 10 
galaxies are available. The shaded regions bracket the 25th and the 75th percentiles of the distributions. 
External mass contributions increase with galaxy stellar mass but decrease with redshift.}\label{fig_msex}
\end{figure*}

We should note that both timescales are calculated using the 
actual (observed) stellar masses of galaxies, not the stellar masses initially formed. 
Because using the latter assigns greater weight to old 
stars and results in earlier formation and assembly times. In practice, however, 
the change affects the two timescales in a similar manner (see Appendix~\ref{ap_tfta} for detailed discussion) and 
thus does not change the overall result. 

\subsection{The contribution of star formation in external galaxies}\label{exms}
Timescale studies shed light on the manner in which galaxies with different masses at different redshifts 
aggregate their stars. But they do not explore quantitatively the roles of internal and external 
processes therein. In this section we evaluate the relative importance of those processes by 
their mass contributions to galaxy assembly. To avoid the mass-loss from stellar evolution, 
we use initial stellar masses in the calculation.
 
For each of our samples, we first trace back along the main branch of its merger tree to identify 
when the main progenitor was involved in a merger event (i.e. mass ratio $\mu\geq1/10$) or 
accretion ($\mu<1/10$) events. We consider the stellar mass of the infalling object 
at the start of the event (i.e. when the merger type is determined) 
to be the mass contribution of that event, under the assumption that all the stars of the 
object will be accreted by the primary host. We sum up the mass that a galaxy has acquired 
from mergers and accretion, and derive the fractional contribution of exteral processes, $f_{ext}$, 
by comparing this mass to the final galaxy mass. Tidally-induced 
shocks and angular momentum loss during a merging process can trigger bursts of star formation, 
contributing to galaxy mass build-up. In our calculation this mass gain is regarded as part of the 
contribution from in-situ star formation.  

Fig.~\ref{fig_msex} shows $f_{ext}$ of low-, intermediate- and high-mass galaxies from 
redshift $z=0$ to $3$. Lines show the median values, while the shaded regions represent 
the 25th to 75th percentiles of the distribution. Both results of the reference 
Ref-L100N1504 (solid lines) and the higher-resolution Recal-L025N0752 (dashed lines) 
simulations are shown in order to demonstrate the convergence of the results. The low-mass 
galaxies at redshift $z=0$ acquire only a small fraction of their mass from external galaxies. 
Over the explored redshift range the median contribution is ${\sim}0.1$ with very little evolution. 
In contrast, galaxies in high-mass bin receive the greatest fractional contribution from 
mergers and accretion in terms of stellar mass gain, with a median of ${\sim}0.19$ and 
a 75th percentile of ${\sim}0.39$. This fraction declines with redshift to ${\sim}0.08$ at $z=2.5$. 
Nevertheless, the low values of $f_{ext}$ for galaxies of any mass at both low- and high-redshifts 
highlight the relative importance of in-situ star formation with respect to external processes 
to the assembly of galaxies.

\subsection{Galaxy merging history}\label{mg}
In preceeding sections, we explored the relative roles that in-situ and external star formation 
play in galaxy mass build-up. In this section, we continue our investigation by 
exploring the seperate contributions of the different external processes in galaxy assembly. 
According to the mass ratio between the two merging systems ($\mu=M_2/M_1$ 
where $M_1>M_2$), these processes are divided into major mergers ($\mu\geq1/4$), 
minor mergers ($1/4>\mu\geq1/10$), and accretion ($\mu<1/10$). 

\subsubsection{Redshift of last major merger}\label{zlst}
\begin{figure} 
\centering\includegraphics[width=8cm,angle=0]{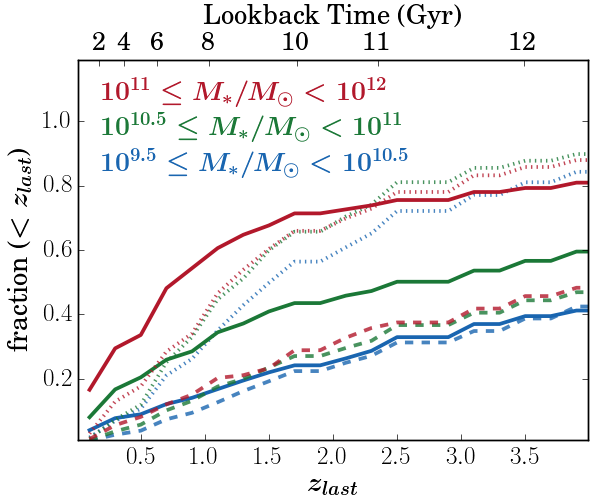}\vspace{-0.cm}
\caption{The cumulative distribution of the redshift of the last major merger event ($\mu\geq 1/4$), $z_{last}$, 
for present-day galaxies ({\it solid lines}) and their parent subhaloes ({\it dashed lines}). 
Galaxies are split into three stellar mass bins as labelled. Only $22\%$ of galaxies with $M_*<10^{10.5}\msun$ have 
experienced a major merger event at $z<1.5$. In contrast, $68\%$ of the most massive galaxies have experienced 
many recent merger events. This mass dependence is not seen in the $z_{last}$ distribution of their parent subhaloes, 
which is due to the non-linear dependence of stellar mass on halo mass.
While the distribution for major subhalo mergers is similar to that of low-mass galaxies, the subhalo $z_{last}$ distribution 
looks more similar to that of high-mass galaxies when minor halo mergers are included ({\it dotted lines}).}\label{fig_zlast}
\end{figure}

Almost all of our present-day galaxies, irrespective of their stellar mass, 
have experienced at least one major merger event in their lives. 
We use the merger trees to determine the redshift, $z_{last}$, when they experienced their
last major merger. Fig.~\ref{fig_zlast} shows the cumulative distribution of $z_{last}$ for 
galaxies in three stellar mass bins. The most massive galaxies have a very active merging history, 
with $68\%$ of the population having been involved in a major merger event since $z=1.5$ 
(a lookback time of $10$~Gyr). This fraction declines with decreasing galaxy mass and 
drops to $41\%$ for intermediate-mass galaxies, and further down to $22\%$ 
for the least massive galaxies. 

Observations of the stellar dynamics of the Milky Way galaxy suggest that no major mergers 
have occurred in the last 10~Gyr \citep{ruchtiEtal15}. Our results show that there is no tension 
between the quiet history of the Milky Way and the CDM paradigm. The Milky Way could easily 
have been drawn from the $\sim 60\%$ of the population that has not undergone a major merger. 
The merger history inferred from the fossil record of the Milky Way is therefore not in conflict with 
those of similar mass galaxies in the EAGLE simulations. 

For comparison, Fig.~\ref{fig_zlast} also show the cumulative distributions of $z_{last}$ for 
the parent subhaloes of those galaxies (dashed lines). Note that we refer to the subhalo mergers as 
the merger events between galaxy-bearing subhaloes. The merger types are determined using the 
same method as for galaxy mergers (see Section~\ref{merger}). In sharp contrast to the active 
merging histories of high-mass galaxies, only $20\%$ of their host subhaloes have undergone a major merger 
event in the last 10~Gyr. Intermediate- and low-mass galaxies share more similarity with their parent subhaloes. 
But, even in the intermediate-mass bin, very recent major mergers between galaxies outnumber those of 
subhaloes by about $10\%-15\%$. This comparison highlights the important difference between the merger 
classification of galaxies and those of subhaloes, especially the massive ones.  
In the high-mass range, the mild dependence of the stellar mass on halo mass 
($M_*\propto M_h^{1/2}$) means that merging galaxies closely matched in mass, 
may have subhaloes of quite different masses. Dotted lines in Fig.~\ref{fig_zlast} show the 
distribution of $z_{last}$ for subhaloes when minor halo mergers are also taken into account. As expected, 
these lines are much more similar to the $z_{last}$ distribution of major mergers of massive galaxies. 

\begin{figure*}  
\centering\includegraphics[width=18cm,angle=0]{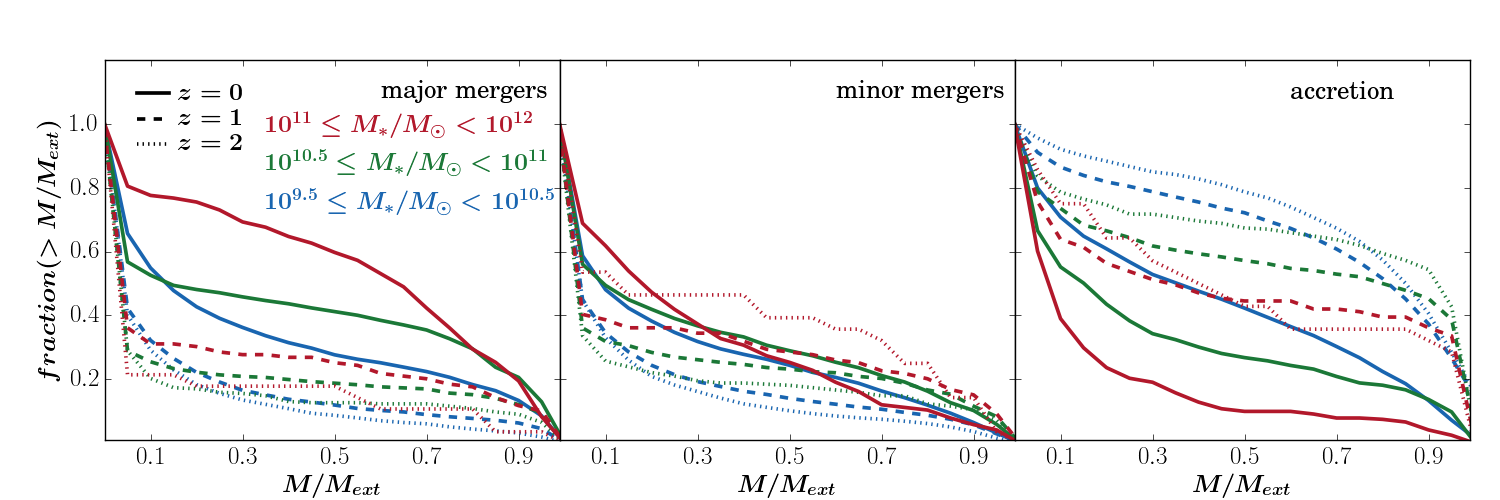}\vspace{-0.cm}
\caption{The fraction of galaxies as a function of the minimum external stellar mass contributed by major mergers ($\mu\geq1/4$, {\it left panel}), 
minor mergers ($1/4>\mu\geq 1/10$, {\it middle panel}) and accretion ($\mu<1/10$, {\it right panel}) at redshift $z=0$ 
({\it solid lines}), $1$ ({\it dashed lines}), and $2$ ({\it dotted lines}). 
The galaxies are split into three bins of stellar mass coded by colours. While accretion plays a larger role than mergers 
in terms of the external stellar mass contribution in low- and intermediate-mass galaxies, at low redshift major mergers  
dominate the external mass contribution in the most massive galaxies.}\label{fig_mgmz}
\end{figure*}

\subsubsection{The contributions of major mergers, minor mergers and accretion}\label{mg_ms}
In this section, we continue our investigation of fractional mass contribution in ~\ref{exms} 
further to explore the respective contributions from major merger, minor merger and accretion 
and their dependence on galaxy mass and redshift. As in Section~\ref{exms}, the initial stellar 
masses of galaxies are used in the calculation in order to remove the impact of stellar evolution 
induced mass-loss.

As the fractional mass contributions show large scatter due to the wide variety of galaxy merging 
histories, we calculate the fraction of galaxies receiving at least a given fractional mass contribution 
from each process. The panels from left to right in Fig.~\ref{fig_mgmz} show the cumulative fraction of galaxies at 
redshift $z=0$ (solid lines), $1$ (dashed lines), and $2$ (dotted lines) as a function of the 
minimum fractional mass contribution from major mergers, minor mergers and accretion, respetively. 
As before, galaxies are binned into low- (blue), intermediate- (green), and high-mass (red) bins. 
Low-mass galaxies at redshift $z=0$ mainly acquire their external masses through accretion, 
while major mergers are the main contributor for their high-mass counterparts. 
Around ${\sim}61\%$ of the most massive population acquired more than half of their external mass 
through major merger events. \cite{parryEF09} arrived at the same conclusion from their analysis of 
semi-analytic models in the Millennium simulation (see Fig.8 in their work). This shift in behaviour is 
driven by the shallow dependence of  stellar mass on halo mass at high halo masses. 
Since $M_* \propto M_h^{1/2}$, a wide range of halo mass ratios lead to mergers occurring 
between galaxies of comparable mass.

Nevertheless, the role of major mergers diminishes with increasing redshift, and at the same 
time, accretion plays a larger role towards higher redshift. As our results show, at redshift $z=2$ 
galaxies of any mass acquired most of their external mass through accretion.

\subsubsection{Evolution of the galaxy merger fraction}
Observationally the frequencies of galaxy pairs and morphologically distorted galaxies 
at different redshifts are commonly used to put constraints on the role of galaxy mergers, 
especially major mergers, in driving galaxy formation. In this section, we examine the census of 
galaxy major mergers, with the aim of shedding light on the evolution of galaxy merger fraction. 

We search galaxy merger trees for galaxies appearing in pairs at each snapshot. 
These pairs are subject to selection criteria somewhat similar to those applied to the 
observational close-pair studies. Any two galaxies are classified as a merging pair if they: 
are separated by a distance $\leq R_{merg}$ ($R_{merg}$ is 5 times the half-stellar mass 
radius of the primary galaxy, see Section~\ref{merger}); have a mass ratio $\mu\geq 1/4$; 
share a common future descendant. The last criterion frees our major merger census from 
the interference of random line-of-sight alignments. 

If two galaxies have not finished merging by $z=0$, 
they will not appear in the same merger tree because they do not have a common descendant. 
As a result, they will not be considered to be a merger pair. \cite{kitzbichlerW08} show that 
the merging times of galaxy pairs can be very extended, leading to a large fraction of pairs surviving to $z=0$. 
To include these pairs in the merger fraction calculation, one should go further in time to 
construct their merging histories after $z=0$. However, as our results show later, neglecting 
the unfinished pairs has only a trivial impact on the global merger fraction.  

We count the number of galaxies that are in pairs. When a galaxy is paired with more 
than one secondary galaxy, the primary galaxy is counted only once. 
The galaxy merger fraction is derived by dividing this number by the total number of galaxies 
at that snapshot. A merger fraction can be converted into a merger rate if we know the merger timescale. 
The time intervals between EAGLE snapshot outputs typically ranges from $0.1\sim1$~Gyr and 
may thus not suffice to derive an accurate estimate of the merger rate. We therefore focus on the 
galaxy merger fraction, rather than the merger rate. Our approach is more readily compared to 
observational measurements (although caution is still warranted because we have not attempted to 
account for observational biases). 

\begin{figure}
\centering\includegraphics[width=8cm,angle=0]{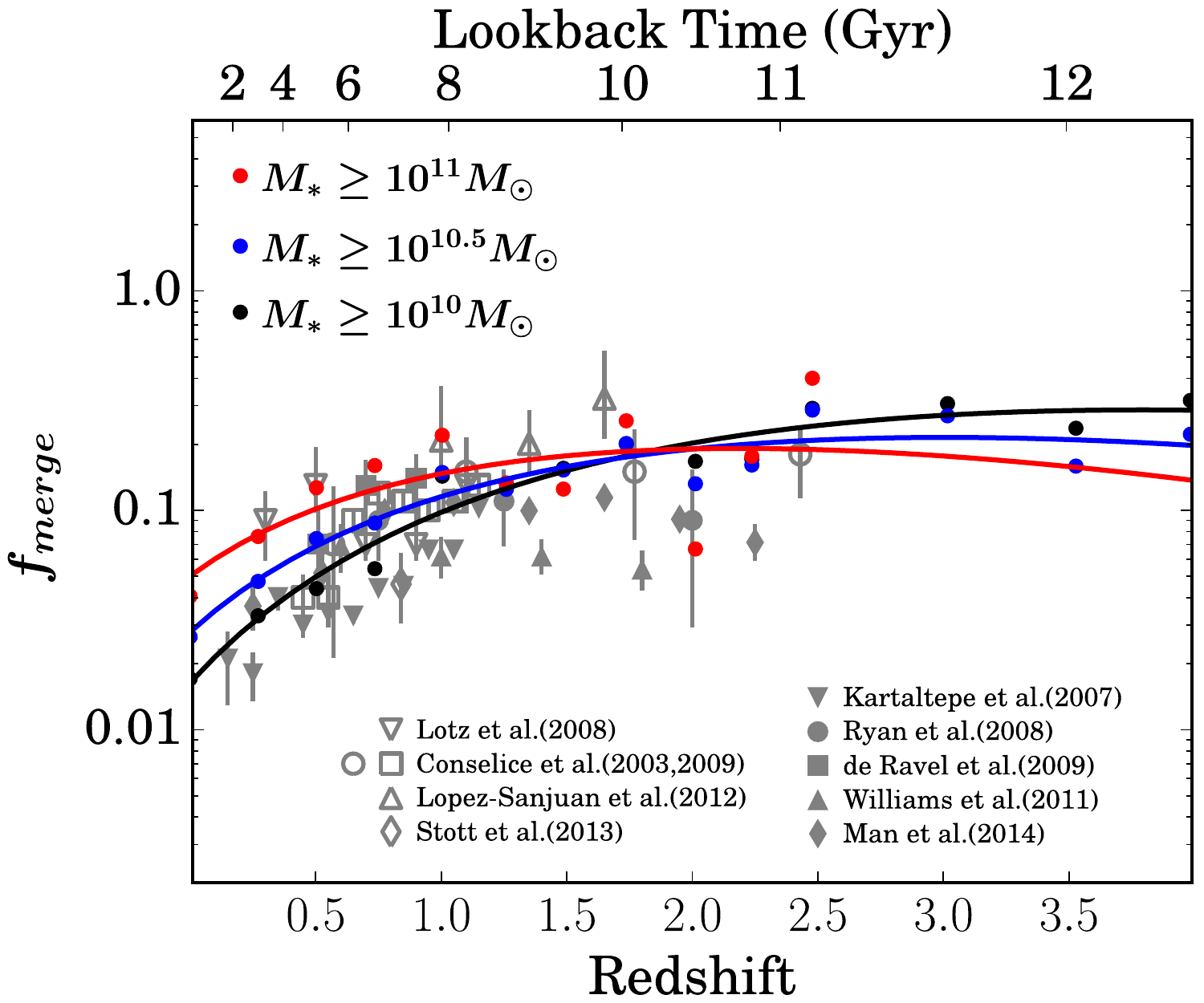}
\caption{Major merger fraction as a function of redshift for galaxies with $M_*\geq 10^{9.5}\msun$ 
({\it black circles}), $\geq 10^{10.5}\msun$ ({\it blue circles}) and $M_*\geq 10^{11}\msun$ ({\it red circles}) 
derived from Ref-L100N1504. The simulation predictions lie within the scatter of the observational data from 
both close-pair studies ({\it solid grey symbols}) and morphological diagnostics ({\it open grey symbols}). 
Curves represent power-law/exponential fits to the simulated merger fraction in the corresponding 
stellar mass bins.}\label{fig_major}
\end{figure}

Fig.~\ref{fig_major} shows the major merger fraction, $f_{\rm merge}$, for galaxies with stellar mass $M_*\geq10^{9.5}\msun$ (black dots), $M_*\geq10^{10.5}\msun$ (blue dots), and 
$M_*\geq10^{11}\msun$ (red dots) over redshift $z=0-4$. The galaxy merger fraction increases monotonically 
towards high redshifts before levelling off at $z\simeq 1-3$, depending on mass. 
The $f_{\rm merge}$of galaxies with $M_*\geq10^{11}\msun$) even declines for $z>2$. 
We compare the simulation predictions with a compilation of real data from both 
galaxy close-pair studies (open symbols) \citep{KartaltepeEtal07,linEtal08,ryanEtal08,deravelEtal09,williamsQF11,manZT14} and galaxy merger studies based on morphological diagnostics (solid symbols) like 
the CAS \citep{conseliceYB09} or the Gini/M20 \citep{conseliceRM08,lotzEtal08,stottEtal13}. 
These data are mostly for galaxies with $M_*\geq10^{10}\msun$. 
Note that this comparison is qualitative since a detailed comparison would require careful 
reconstruction of the observational criteria. Overall, however, the predicted galaxy merger fraction 
lies within the scatter of observational data, but is most compatible with studies based on 
morphological analysis. 

Observational studies often parametrize the redshift-dependence of the galaxy merger fraction 
as a power law, $\propto(1+z)^n$, with index $n=0-4$. However, the rise of the merger fraction 
beyond redshift $z\approx1$ is not as rapid as it is at $z<1$, especially for massive galaxies. 
\citet{conseliceYB09} show that a combined power-law/exponential function can fit both 
the steep increase of the observed merger fraction at $z\sim0-1$ and the plateau beyond. 
We use a combined fitting function $a(1+z)^b e^{c(1+z)}$ to fit the simulation predictions, 
in which $a$, $b$, $c$ are free parameters and $z$ is the redshift. The curves in Fig.~\ref{fig_major} 
represent the least-square fitting results in three mass bins. Table~\ref{tab_fit} lists the best-fit values of the 
three parameters and their 1-$\sigma$ uncertainties obtained from the fitting. 

\begin{table}
\caption{The values of the parameters $a$, $b$, $c$, with 1-$\sigma$ uncertainties, of  
a power-law/exponential fitting function $a(1+z)^b e^{c(1+z)}$ in which $z$ is redshift. 
These values are determined by the least-square fittings to the predicted galaxy merger 
fraction in three stellar mass bins.}\label{tab_fit}
\begin{tabular}{@{}lccc@{}} \hline
$M_*/\msun$ & a & b & c \\ \hline
$\geq 10^{10}$   & $0.035\pm0.069$ & $3.694\pm0.519$ & $0.771\pm0.206$ \\
$\geq 10^{10.5}$ & $0.062\pm0.074$ & $3.206\pm0.560$ & $0.801\pm0.222$\\
$\geq 10^{11}$    & $0.122\pm0.422$ & $2.833\pm2.433$ & $0.889\pm1.195$ \\ 
\hline\end{tabular}
\end{table}

The merger diagnostics are also sensitive to merger mass ratios, we also consider the impact on our results of 
extending the merger mass ratio to a smaller value ($\mu\geq 1/10$). We find that the merger fraction is 
elevated by only a factor of $1.5-1.8$ on average by the inclusion of minor merging events, and shows 
similar trends with redshift. 

\subsection{The impact of feedback on galaxy mass assembly}\label{fb}
\begin{table*} 
\caption{Values of the parameters used in the simulations with varying feedback efficiency:
size of the simulation volume ($L$), particle number ($N$), dark matter and initial baryonic particle mass ($M_{DM}$ and $M_g$), 
the asymptotic minimum and maximum value of stellar feedback efficiency ($f_{th, min}$ and $f_{th, max}$),  
accretion disc viscosity $C_{visc}$, and the temperature increment of stochastic AGN heating ($\Delta T_{AGN}$). 
We refer readers to \protect\cite{crainEtal15} for detailed information on these paramertes.}
\begin{tabular}{lcccccccc} \hline
 Identifier & $L$ & $N$ & $M_{DM}$  & $M_g$ & $f_{th, min}$ & $f_{th, max}$ & $C_{visc}$ & $\Delta T_{AGN}$ \\
  & $[cMpc]$ & & [$M_{\odot}$] & [$M_{\odot}$] & & & & [$K$] \\ \hline
Ref-L100N1504 &$100$& $1054^3$ & $9.70\times 10^6$ & $1.81\times 10^6$ & $0.3$ & $3.0$ & $2\pi$ & $10^{8.5}$ \\ 
StrongFB& $50$ & $752^3$ & $9.70\times 10^6$ & $1.81\times10^6$ & $0.6$& $6.0$ & $2\pi$ & $10^{8.5}$\\
WeakFB  & $25$ & $375^3$ & $9.70\times 10^6$ & $1.81\times10^6$ & $0.15$  & $1.5$ & $2\pi$ & $10^{8.5}$\\
NoAGN  & $25$ & $375^3$ & $9.70\times 10^6$ & $1.81\times10^6$ & $0.3$  & $3.0$ & - & - \\ \hline
\end{tabular}\label{tab_fb}
\end{table*}

So far, we have shown that the assembly of massive galaxies is very different 
to that of their smaller counterparts. A very interesting question is whether this is due to the 
feedback from star formation and black hole growth. AGN feedback, for example, is able to 
efficiently suppress in-situ star formation by heating the hot coronae
of galaxies and suppressing the inflow of cool gas \citep{bowerEtal16}. 
To gain more insight on this aspect, we calculate the mass contribution of internal and 
external processes in simulations with varying efficiencies of feedback from stars and AGN. 
These runs differ in simulation volume but have the same resolution. Table~\ref{tab_fb} 
lists the values of the parameters used in their feedback models. The effect of these
changes on the stellar mass function and galaxy star formation rates is considered in \cite{crainEtal15} .

\begin{figure*}  
\centering\includegraphics[width=18cm,angle=0]{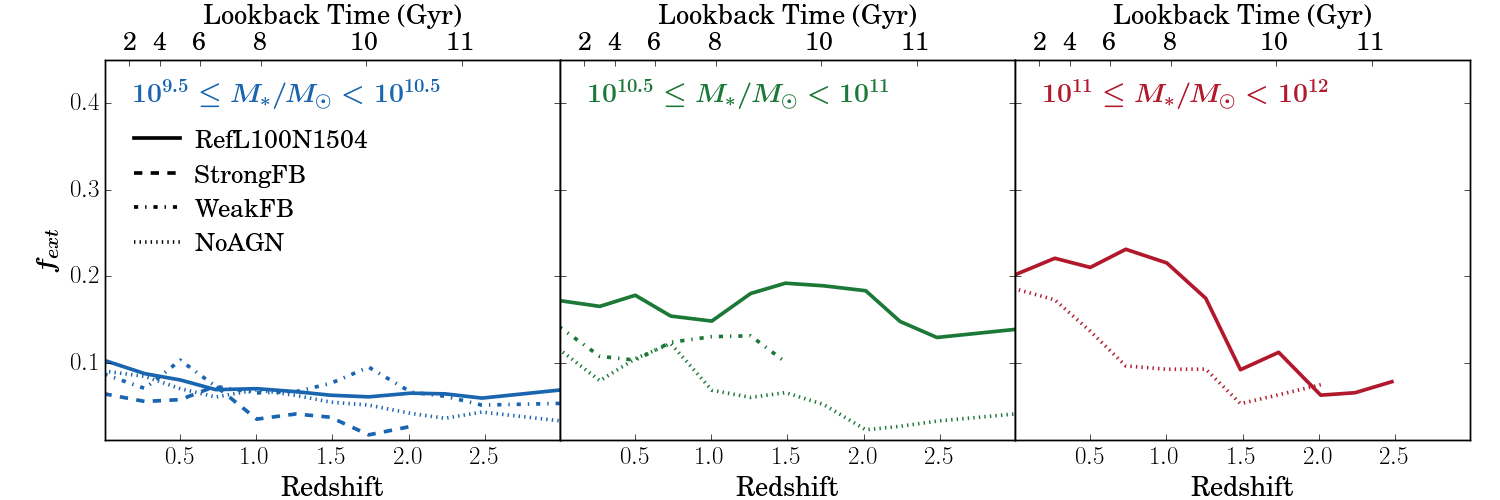}\vspace{-0.cm}
\caption{The fractional mass contribution of mergers and accretion for galaxies at redshifts $z=0-3$ when there is  
strong ({\it dashed lines}) and weak ({\it dash-dotted}) stellar feedback, and no AGN feedback ({\it dotted lines}). 
The reference model (solid lines) is also shown for comparison. We split the galaxies into three stellar 
mass bins. We only show points for which more than 10 galaxies contribute. Changes in the efficiency of star 
formation and the role of AGN make significant differences to the external stellar mass fraction.}\label{fig_fb}
\end{figure*}

The panels from left to right in Fig.~\ref{fig_fb} compare the fractional mass contribution from mergers and accretion, 
$f_{ext}$, for low-, intermediate- and high-mass galaxies over redshift $z=0$ to $3$ in the presence of  
weak (dot-dashed lines) and strong (dashed lines) stellar feedback, and no AGN feedback (dotted lines). 
The results for the reference model (solid lines) are also shown for comparison. The lines show the 
median of the $f_{ext}$ distribution and stop at the redshift when fewer than 10 galaxies are available for analysis. 
We find that stellar feedback has very limited impact on the mass build-up of low-mass galaxies (left panel). 
Increasing or decreasing the feedback efficiency leads to only $\lesssim 5\%$ of changes in their $f_{ext}$. 
In the strong feedback case, the analysis consistently suggests a slight decrease of  $f_{ext}$ as more of the 
star-forming gas within small galaxies is lost in outflows, reducing their contribution to the stellar mass. 
The formation of massive galaxies is also strongly suppressed, however, and the small simulation volume ($25$ cMpc) 
prevents us reliably determining if there is an increase in $f_{ext}$ in the few large objects that form. 
In the case of weak stellar feedback, the efficiency of galaxy formation is similarly  increased over a wider range  
of halo mass, and $f_{ext}$ changes little in the left panel.  In the middle panel, $f_{ext}$ is lower than 
the reference simulation (and is more similar to the curve in the left panel). In the absence of effective 
stellar feedback, AGN feedback has a similar impact in high and low mass haloes \citep{bowerEtal16} and 
we expect the differences between the panels to be smaller, as seen. 

Galaxies in the first two panels are insensitive to the AGN feedback since (in the reference model) 
star formation driven outflows oppose the build-up of high gas densities in the central regions \citep{bowerEtal16}. 
In contrast, the AGN feedback has a very noticeable impact on the $f_{ext}$ of their massive counterparts. $f_{ext}$ 
declines in the absence of AGN feedback, consistent with the negative impact of AGN feedback on in-situ 
star forming in massive galaxies. This explains many of the differences, but not all of them. For example, 
for the most massive galaxies, there is still a rapid rise in $f_{ext}$ to the present day that may be related to 
the recent cosmological acceleration of the Universe. 

\section{Comparisons to other work}\label{dis}
In this work we focus on the assembly and formation of galaxies. This is a topic that has been extensively 
studied using N-body simulations and semi-analytic galaxy formation models. 

\cite{kauffmannCW96} and \cite{deluciaEtal06} already show that the formation time of brightest cluster 
galaxies is much earlier than their assembly time and \cite{parryEF09} show that, with the exception of the 
brightest galaxies, major mergers are not the primary mechanism by which most galaxies assemble their mass. 
Our hydrodynamic EAGLE simulations exhibit the same trends and their dynamic range allows us to contrast 
the formation of the most massive galaxies with that of galaxies similar to the Milky Way. We do not, 
however, find galaxies with formation and assembly times as large as in \cite{deluciaB07}, who find that 
galaxies with $M_*>10^{12}{\rm M}_\odot$ form 50\% of their stars at $z\approx5$; galaxies in our 
highest-mass bin only cover the range $10^{11}-10^{12} {\rm M}_\odot$ and form most of their stars at $z>2$. 

\cite{guoW08} compare the contributions from star formation and galaxy mergers to the mass 
build-up of galaxies using semi-analytical models. In common with our results, they find that 
major merger play an important role in the growth of galaxies more massive than the Milky Way 
and that the relative importance of star formation increases towards high redshift. Nevertheless, 
we disagree with their conclusion about major mergers also dominating the growth of high-redshift 
massive galaxies. Our results show that in-situ star formation, instead of major mergers, is the 
dominant contributor to those galaxies. 

\cite{lacknerEtal12} examine galaxy formation and assembly histories in adaptive mesh refinement (AMR) 
simulations. Their results show that the accreted fraction has a smooth dependence on stellar mass, 
but their calculations do not include AGN feedback and do not capture the observed break in the 
galaxy stellar mass function. The importance of feedback can be recognized by looking at zoom simulations of 
galaxies similar to the Milky Way. Our results disagree with the high accreted star contributions 
reported by \cite{oserEtal10}. This discrepancy is presumably due to the lack of effective feedback at high 
redshift in their runs, as the in-situ fraction can be drastically reduced in simulations without any 
feedback \citep{hirschmannEtal12}. 

\cite{rodriguezgomezEtal16} also provide some insight on this topic 
using the Illustris simulation \citep{vogelsbergerEtal14}. Similar to our results, they confirm 
the greater role of mergers and accretion in the mass growth of present-day 
massive galaxies with $M_*>10^{11}{\rm M}_\odot$ as well as the decreasing importance of these processes 
with increasing redshift. This similarity supports the robustness of our conclusions to varying subgrid physical models. 
In particular, we make a comparison between the results with varying feedback efficiencies,  
shedding light on how stellar and AGN feedback affect the mass build-up of galaxies. 
Nevertheless, there is also disagreement between our results and that of \cite{rodriguezgomezEtal16}. They highlight 
the importance of major mergers in contributing to the assembly of low-mass galaxies and high-mass galaxies alike. 
This contrasts with our results which show that accretion, rather than major mergers, are the main contributors in 
low-mass galaxies. This discrepancy may be the result of the different methods the two works used for merger type 
determination.  While \cite{rodriguezgomezEtal16} define the type of a merger event when the secondary galaxy 
reaches its maximum stellar mass, we determine the merger type when the secondary galaxy is some distance, 
5 times the half-stellar mass radius of the primary galaxy, from the primary host. Unfortunately, the time interval of 
the EAGLE snapshot outputs is not sufficient to demonstrate that the two methods look at the same merging epoch and 
classify the mergers in the same way. It is also worth noting that the mass-loss from stellar evolution is taken into account 
in our mass contribution calculation and that Illustris simulation has a steeper slope to the faint-end galaxy mass function, 
compared to our simulation and observations. The most fundamental difference may, however, 
be the implementation of stellar and AGN feedback. These are very different in the simulations, and we have shown that this can lead to significant differences in galaxy assembly histories in \S3.5. Clearly this is an interesting avenue for more detailed future investigation.

\cite{mosterNW13} and \cite{behrooziWC13} consider the topic from a more observational perspective, 
using the abundance matching method. They find increasing trends in the fraction of accreted stars with 
increasing galaxy mass and decreasing redshift that agree closely with 
our simulations. The consistency between the empirical results and the simulation predictions provides 
encouraging support both to our results and to the EAGLE simulation runs.

Our results show that the mass assembly of galaxies, however, is not simply a reflection of the growth of their parent 
haloes. Additional physical processes, such as stellar and AGN feedback, make galaxy formation 
efficiency a strong function of halo mass $M_h$. The resulting stellar mass-halo mass relation has a 
steep slope in low-mass haloes ($M_*\propto M_h^{2}$) and a shallower slope at high mass 
($\propto M_h^{1/2}$) \citep[e.g.][]{bensonEtal03}. The steep low-mass slope arises because 
the binding energy per unit mass of the halo scales with the halo mass as $M_h^{5/3}$, while the energy 
available from stars is proportional to the stellar mass. The high-mass slope arises because 
AGN feedback is able to suppress star formation and because the cooling time is long in 
massive haloes \citep[e.g.][]{ReesO77, SilkR98, bensonEtal03,bowerEtal06, crotonEtal06}, 
leaving galaxy merging as the only effective growth channel. From an observational perspective, 
the connection can be derived by matching the abundance of galaxies and haloes assuming a monotonic 
relation \citep[e.g.][]{kravtsovEtal04}. The dark matter mass function is described as 
a power-law (with index $\sim-1$) with only a slow rollover at high mass. In contrast, 
the galaxy stellar mass function is almost flat at low mass \citep[e.g.][]{fontanaEtal06}. 
Matching the two by abundance requires a quadratic dependence of stellar mass on halo mass. 
At high mass, the galaxy stellar mass function has a sharp break implying that haloes of 
increasing mass host galaxies of very similar mass. 

The discrepancy implied by these transformations makes mapping the merger histories of 
haloes to those of galaxies non-trivial and halo mass dependent \citep[e.g.][]{cattaneoEtal11}. 
For low-mass galaxies and haloes, a halo merger with a mass ratio $1/4$ may correspond (roughly) 
to a merger between galaxies of mass $\sim1/16$. For massive galaxies, a minor halo merger 
(between a massive halo and a satellite halo) may actually correspond to a major (almost equal-mass) 
galaxy merger. In this high-mass regime, assuming a uniform galaxy formation efficiency to derive 
galaxy merging histories from halo merging histories inevitably 
underestimates the importance of major galaxy mergers, and overstates the role of minor mergers. 
Many papers have pointed out the disagreement between the galaxy merger rate and the halo merger rate 
\citep[e.g.][]{berrierEtal06, parryEF09, hopkinsEtal10,guoEtal11}, and here we are able to 
demonstrate this directly. We compare the times (in redshift, $z_{last}$) when galaxies and 
their parent subhaloes experience their last major merger events, and find that the $z_{last}$ 
distribution of massive galaxies differs greatly from that of their host subhaloes. The former looks 
more closely like the $z_{last}$ distribution of subhaloes only when minor subhalo mergers are 
also included.

The principal aim of this paper has been to quantify the role of mergers in the formation histories of galaxies 
in the EAGLE reference simulation. Since the simulation provides a good description of the galaxy stellar mass function and 
its evolution, as well as many other aspects of the observable Universe, we make the implicit assumption that 
the formation histories of the simulated galaxies provide a good approximation to those of galaxies in the real Universe. 
The long timescales of galaxy evolution make it impossible to observe the growth of galaxies directly; nevertheless, 
it may be possible to reconstruct the build up of one galaxy, the Milky Way, from careful archaeology of its stellar content and 
their the use of chemical tagging techniques \citep{HoggEtal16}. Unfortunately, the formation history of galaxies like 
the Milky Way is extremely diverse, and careful thought will be required to understand how results, such as those 
from the GAIA satellite, can be used to reach definitive conclusions.

\section{Summary and conclusions}\label{con}
In this paper, we have investigated the assembly and merging histories of hundreds of thousands of 
central galaxies in the EAGLE cosmological simulation project. The hydrodynamic simulations
include a range of gas, stellar and black hole physical processes relevant to galaxy formation, and 
have been shown to match the properties of observed galaxies reasonably well. Because of this, 
these simulations provide an ideal test bed for elucidating the roles 
played by galaxy mergers and in-situ star formation in galaxy formation.

We construct galaxy merger trees by applying the D-Trees algorithm \citep{jiangEtal14} to SUBFIND 
subhalo catalogues across snapshot outputs. They enable us to chronicle galaxy formation from 
$z=3$ to the present day. Because galaxies will slowly lose stellar mass due to tidal stripping before 
they finally merge, a careful definition of the masses of galaxies prior to and during a merger is required. 
In this paper we use a definition based on a separation of 5 times the 
galaxy half-stellar mass radius to signal the start of a merging event and then determine the merger type. 
According to the mass ratio between the primary and the secondary galaxies, merger events are 
classified as either major mergers (with mass ratios $\mu\geq1/4$), minor mergers ($1/4>\mu\geq1/10$), 
or accretion ($\mu<1/10$). Considering that galaxies also suffer mass-loss due to stellar evolution, 
we use the initial stellar mass, i.e. the stellar mass being formed, when evaluating the relative contributions of 
in-situ and external processes to the mass growth of galaxies.

Our main results are summarised as follows:
\begin{itemize}
\item We contrast the assembly time ($t_a$, when the main progenitor of a galaxy had assembled 
half its present-day stellar mass) and the formation time ($t_f$, when that mass had formed, 
regardless in which progenitor) of galaxies. Galaxies less massive than $10^{10.5}\msun$ 
have very similar $t_f$ and $t_a$, showing that most of their stars formed in their main progenitors. 
Above a mass of $10^{10.5}\msun$, galaxies are dominated by increasingly old stars, 
but for the most massive galaxies the assembly time decreases, implying that although the stars are old, 
they have only recently been assembled into the present-day galaxies (Fig.~\ref{fig_tfta}). 
We also compare the formation and assembly times of galaxies with those of their parent subhaloes and 
find quite different trends. The $t_f$ and $t_a$ of the subhaloes, in contrast, show a high level of similarity 
over the mass range studied, decreasing monotonically with increasing mass (Fig.~\ref{fig_h_tfta}). 

\item We quantify the mass fraction of stars that are formed `in-situ\rq{} vs. stars that have an 
accreted origin. Galaxies less massive than $10^{10.5}\msun$ typically acquire less than $10\%$ of 
their mass through galaxy mergers or accretion of stars formed in other systems. In contrast, in galaxies 
more massive than $10^{11}\msun$, typically ${\sim}20\%$ of the system's stars have an external origin. 
There is considerable scatter in both cases (Fig.~\ref{fig_msex}). 

\item The fraction of accreted stellar mass in less massive galaxies evolves mildly with 
 redshift. In the high-mass galaxies, the assembly and formation times become increasingly 
 similar with increasing redshift and the fraction of externally formed stellar mass declines 
 (Fig.~\ref{fig_dtmz} and \ref{fig_msex}).
 
\item We measure the distribution of the redshifts when galaxies have their last major mergers. 
For galaxies less massive than the Milky Way, the median redshift of the last major merger 
is $z\approx2$, which is compatible with the quiet formation history of the Milky Way 
implied by recent observations (Fig.~\ref{fig_zlast}). 

\item Accretion dominates the external mass contribution for less massive galaxies, 
while major mergers become the main mass contributor of external mass for massive galaxies 
(Fig.~\ref{fig_mgmz}).

\item We compute the fraction of galaxies in a snapshot that are undergoing 
major mergers, and explore the variation of this fraction with redshift and galaxy mass. We find that 
the merger fraction rises rapidly between the redshifts $z=0$ and $1$, but flattens at higher redshift. 
Given the uncertainties inherent in the comparison, and the range of methods applied to observational 
datasets to this diagnostic, our simulation predictions display a remarkable similarity with 
observational studies (Fig.~\ref{fig_major}).

\item Strengthening {\it or} weakening stellar feedback results in a decline in the external mass contribution to galaxies. 
While low-mass galaxies are weakly affected by AGN feedback, their massive counterparts show 
a significant reduction in the external mass contribution (Fig.~\ref{fig_fb}). These changes can be broadly \
understood as resulting from changes in the efficiency of on-going star formation and the impact of AGN feedback. 
\end{itemize}

Overall, we find general agreement between our results and studies based 
on semi-analytic models. Massive galaxies are found to have started their star formation 
earlier than low-mass galaxies but partly in objects other than the main progenitor, 
and then assembled those stars later through mergers and accretion. 
This assembly history also implies that they have older stellar populations, 
consistent with the `downsizing\rq{} trend seen in many observational studies. 

Despite the close relationship between galaxies and their parent haloes, their formation and assembly 
histories are very different. The formation of dark matter haloes, contrary to that of galaxies, is typically 
hierarchical, in the sense that their formation times decrease with increasing halo mass. The assembly of 
haloes also proceeds in a different manner from that of the galaxies that they host. Massive galaxies 
acquire a fractional mass from major and minor mergers, while their parent haloes grow in mass 
mainly by smooth accretion. 
 
As in \cite{guoW08}, we compare the stellar mass contributions from in-situ star formation and 
external processes to galaxies of various stellar masses and redshifts. These comparisons highlight 
in-situ star formation as the main mechanism in the formation of low- and high-mass galaxies alike 
at both low- and high-redshifts. Our investigation also confirms the role of mergers and accretion 
in the formation of massive galaxies, which has been revealed by many semi-analytic studies 
\citep[e.g.][]{deluciaEtal06,deluciaB07,guoW08}. This result can be attributed to the supermassive 
black holes developed in these galaxies. Their energetic feedback prevent gas from cooling down to 
form stars. Massive galaxies, as a result, have no other ways to grow in mass but to accrete stars from other systems. 
Among all external processes, major mergers contribute most to the addition of external mass to 
present-day massive galaxies, while accretion is the main contributor for their 
less massive counterparts. At higher redshift, accretion dominates the external mass contribution for 
galaxies of any mass. We find both agreements and discrepancies between our results and those of other recent 
simulations. Some of the discrepancies may result from the different ways in which mergers are identified and 
their mass contribution are evaluated.  

The galaxy merger trees that we construct, and the role of galaxy mergers that we quantify here, will 
also be used in future work looking at other aspects of the galaxy population. For example, the dependence of 
galaxy size on merger history is considered as part of \cite{furlongEtal15b}, and their role in driving colour 
evolution is considered in \cite{trayfordEtal16}. These works are focused on the observational aspect, 
while this paper is focussed on the underlying physical process. The results we present, of course, 
come with the caveat that the simulation is not {\it the} Universe, and must be understood as applying to 
an approximation of reality. With future observational facilities, it may become possible to test the results 
we present directly.

\section*{Acknowledgement}
We acknowledge our colleagues for very useful comments and suggestions in the writing of this paper. 
RGB thanks Gary Mamon for very insightful discussion. RAC is a Royal Society University Research Fellow. 
YQ acknowledges the support of STFC Grant ST/L00075X/1. This work used the DiRAC Data Centric system at 
Durham University, operated by the Institute for Computational Cosmology on behalf of the STFC DiRAC HPC 
Facility (www.dirac.ac.uk). This equipment was funded by BIS National E-infrastructure capital grant ST/K00042X/1, 
STFC capital grant ST/H008519/1, and STFC DiRAC Operations grant ST/K003267/1 and Durham University. 
DiRAC is part of the National E-Infrastructure. We also acknowledge 
the PRACE for awarding us access to the resource Curie based in France at Tr\`{e}s Grand Centre de Calcul and 
the support from Interuniversity Attraction Poles Programme initiated by the Belgian Science 
Policy Office ([AP P7/08 CHARM]). This work is sponsored in part by the European Research Council 
(grants GA 267291 and 278594-GasAroundGalaxies) and 
by the STFC ``Consolidated Grant'' to Durham University (ST/F001166/1).

\bibliographystyle{mnras}
\bibliography{revised_ms}

\appendix
\section{Separation criteria}\label{ap_r}
\begin{figure}  
\centering\includegraphics[width=8cm,angle=0]{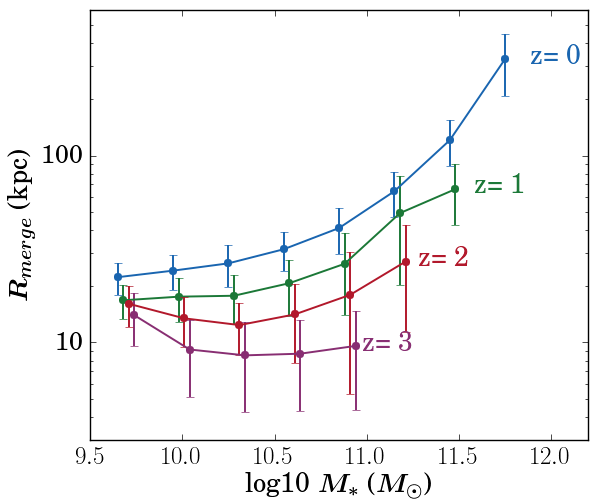}\vspace{-0.cm}
\caption{The characteristic physical separation, $R_{merge}$, for merging galaxies over 
the stellar mass and redshift ranges explored in this work. The solid lines represent the means of 
the distributions, while the error bars depict the 25th and the 75th percentiles. 
$R_{merge}$ is defined as five times the half-stellar mass radius of the primary galaxy. The type of 
a galaxy merger is determined when the two merging galaxies are not less than $R_{merge}$ apart.}\label{ap_sep}
\end{figure}

The type of a galaxy merger event is determined by the mass ratio between the secondary and the primary 
galaxies when they are separated by some minimum distance. We do this so that the secondary galaxy is not 
strongly affected by tidal-induced mass loss. The separation criterion, $R_{merge}$, is defined as 
$R_{merge}=5\times R_{1/2}$ where $R_{1/2}$ is the half-stellar mass radius of the primary galaxy. 
Note that $R_{merge}$ is a 3D separation. Fig.~\ref{ap_sep} illustrates the $R_{merge}$ distribution 
as a function of the galaxy stellar mass at redshifts $z=0$, $1$, $2$, and $3$. These ranges of values are 
in roughly accord with the projected separations used in observational galaxy pair studies.

\section{Impact of mass-loss on the formation and assembly times}\label{ap_tfta}
\begin{figure}
\centering\includegraphics[width=8cm,angle=0]{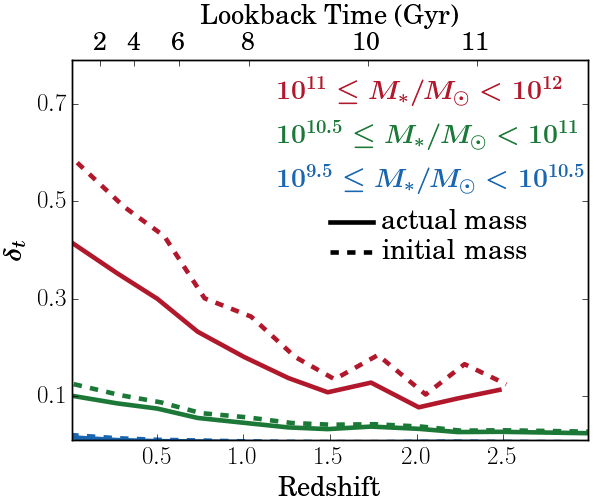}\vspace{-0.cm}
\caption{The median of $\delta_t$ as a function of redshift for galaxies in different mass bins 
(indicated by colours and legends). The $\delta_t$ is calculated using either initial stellar mass 
({\it dashed lines}) or actual stellar mass ({\it solid lines}). Using initial stellar mass leads to 
a quantitative change in $\delta_t$ at fixed redshift, especially for massive galaxies, but does not 
change its evolutionary trend with redshift.}\label{ap_fig_dt}
\end{figure}

\begin{figure*}
\centering\includegraphics[width=10cm,angle=0]{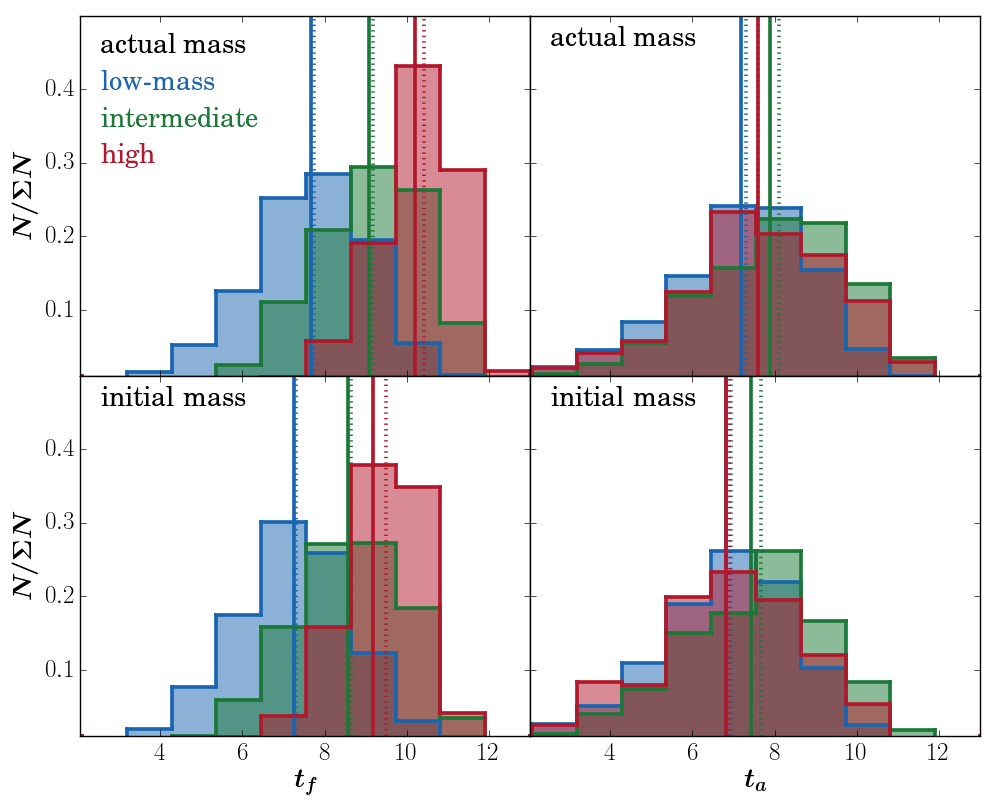}\vspace{-0.cm}
\caption{Comparisons of the formation time, $t_f$, and the assembly time, $t_a$, that are calculated 
using actual stellar mass ({\it top panels}) and initial stellar mass ({\it bottom panels}) for galaxies at $z=0$. 
Galaxies are split into three stellar mass bins as indicated by colours and legends. Vertical solid lines indicate 
the average values of the distributions while the dotted lines the medians.}\label{ap_fig_tfta}
\end{figure*}

We use the actual stellar mass (i.e. the stellar mass observed) of a galaxy to define its formation time, 
$t_f$, and its assembly time, $t_a$. However, galaxies continuously experience mass-loss due to 
stellar evolution during their lifetimes \citep[see Fig.1 in][]{SegersEtal16}. Neglecting this mass-loss 
in the timescale calculation would inevitably lead us to an earlier epoch (corresponding to a larger 
lookback time) to define the $t_f$ and $t_a$. To address the impact, we calculate again the $t_f$ and 
the $t_a$ of our sample galaxies but using the initial stellar mass (i.e. the mass initially formed). 
Fig.~\ref{ap_fig_tfta} compares the distributions of $t_f$ and $t_a$ calculated using actual stellar 
mass in three galaxy mass bins ({\it top panels}) to those based on initial stellar mass ({\it bottom panels}). 
As expected, the galaxies have a relatively smaller $t_f$ and $t_a$ when initial stellar mass is used. 
This change occurs in a similar manner for low-mass galaxies and massive galaxies. As a result, 
the relative difference between the timescales, $\delta_t$, shows a similar trend with redshift as 
that of actual stellar mass, as shown in Fig.~\ref{ap_fig_dt}. Taking into account the mass-loss 
from stellar evolution would therefore not change our conclusions.

\section{Effect of the aperture}\label{ap_ap}
\begin{figure*}  
\centering\includegraphics[width=18cm,angle=0]{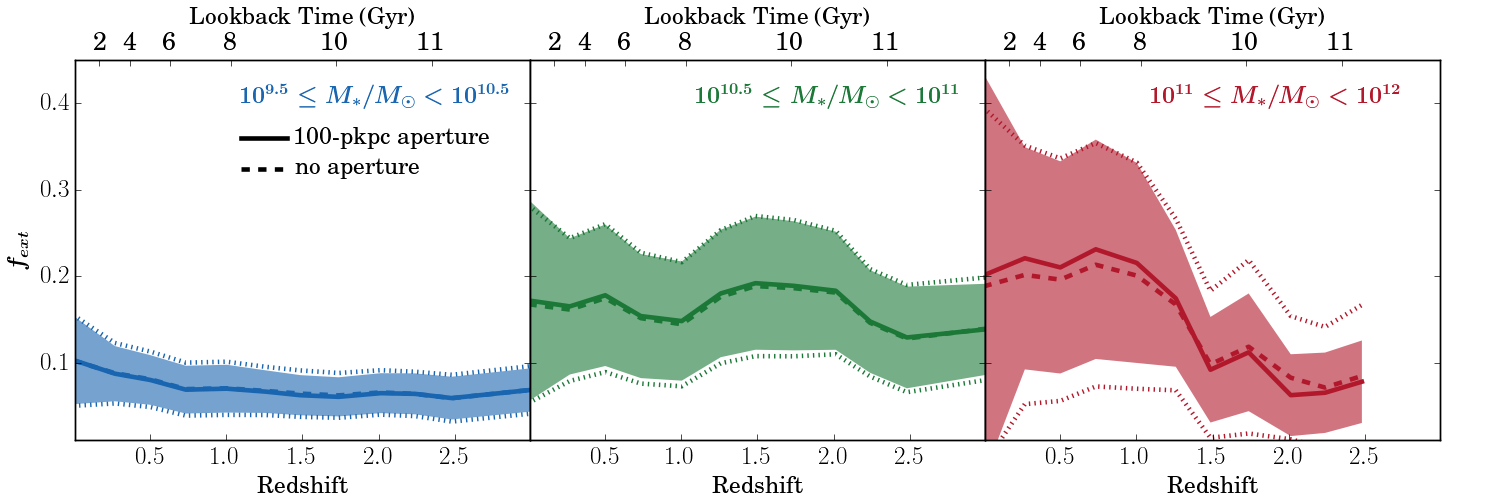}\vspace{-0.cm}
\caption{The fractional mass contribution of mergers and accretion with ({\it solid lines}) and 
without ({\it dashed lines}) a 100 pkpc aperture for galaxies at redshifts $z=0-3$. The galaxies 
have been split into three stellar mass bins as labelled. Lines represent the medians of the 
distributions while the shaded regions ({\it dotted lines}) mark the 25th and the 75th 
percentiles.}\label{ap_exms}
\end{figure*}

In this work the galaxy mass is defined as the actual (or initial) stellar mass enclosed by a spherical 
aperture with a galactocentric radius of $100$~pkpc (proper kpc). The use of an aperture enables us to 
focus on the central part of a galaxy where the main properties are derived. Nevertheless, as shown by 
\cite{schayeEtal15}, this aperture choice will have an impact on the mass measurement of massive 
galaxies ($\geq 10^{11}\msun$). Compared to their less massive counterparts, massive galaxies 
usually experience more merging events. Some of their stars may be deposited in the outer regions 
by the tidal force during a merging process, forming a diffuse and faint intracluster light (ICL) as 
observed in the centre of galaxy clusters \citep[e.g.][]{theunsW97, behrooziWC13}. The galaxy mass 
can be underestimated when an aperture is employed in the mass measurement. 
For example, we compare the present-day galaxy masses measured using a $100$~pkpc aperture to 
those derived by the SUBFIND algorithm (without an aperture). 
For low- and intermediate-mass galaxies, there is no difference between the two masses. 
But for the most massive galaxies, we find that ${\sim}30\%$ of the stellar mass lies 
outside of the aperture. 

To evaluate the mass contributions of external processes to the growth of a galaxy, 
we sum up the stellar mass that the galaxy has acquired from mergers and accretion 
and compare it to its final stellar mass. By using an aperture mass we may underestimate the total stellar 
mass of a massive galaxy, and thus overestimate the fractional mass contributions of external processes, $f_{ext}$. 
Fig.~\ref{ap_exms} compares the distribution of $f_{ext}$ based on a $100$~pkpc aperture (solid lines) 
and no aperture (dashed lines) for galaxies in three stellar mass bins at redshifts $z=0-3$. 
Lines represent the medians of the distributions. The shaded regions and dotted lines depict 
the 25th and the 75th percentiles of the distributions in the $100$~pkpc aperture case and no-aperture 
case, respectively. Using $100$~pkpc aperture masses has almost no impact on our results when galaxies are 
less massive than $\leq10^{11}\msun$, and the impact remains small for even more massive galaxies.

\end{document}